\documentstyle[a4,12pt,epsfig,amsmath]{article}
 \def\relic{\Omega_{DM
}}
\def\lsim{\raise0.3ex\hbox{$\;<$\kern-0.75em\raise-1.1ex\hbox{$\sim\;$}}}
\def\gsim{\raise0.3ex\hbox{$\;>$\kern-0.75em\raise-1.1ex\hbox{$\sim\;$}}}
\def\neut{\tilde\chi_1^0}
\def\higgsu{m_{H_2}^2}
\def\higgsd{m_{H_1}^2}

\def\neumass{m_{\tilde\chi_1^0}}

\def\higgsu{m_{H_u}^2}
\def\higgsd{m_{H_d}^2}
\def\higgsuew{m_{H_u}^2}
\def\higgsdew{m_{H_d}^2}

\def\neumass{m_{\tilde\chi_1^0}}

\def\neut{\tilde\chi_1^0}
\def\bsg{$b\to s\gamma$}

\def\asusy{a^{\rm SUSY}_\mu}

\DeclareMathAlphabet   {\mathsc}{OT1}{cmr}{m}{sc} 
 
\def\[{\left [} 
\def\]{\right ]} 
\def\({\left (} 
\def\){\right )}

\newcommand{\gappeq}{\mathrel{\rlap {\raise.5ex\hbox{$>$}} 
{\lower.5ex\hbox{$\sim$}}}} 
\newcommand{\lappeq}{\mathrel{\rlap{\raise.5ex\hbox{$<$}} 
{\lower.5ex\hbox{$\sim$}}}}

\newcommand{\bea}{\begin{eqnarray}}
\newcommand{\eea}{\end{eqnarray}}

\begin{document}

\pagestyle{empty}


\rightline{FTUAM 04/10}
\rightline{IFT-UAM/CSIC-04-25}
\rightline{July 2004}

\vspace{1.cm}

\begin{center}

{\Large {\bf Gamma-ray detection from neutralino annihilation\\
in non--universal SUGRA scenarios}}
\vspace{1cm}\\
{\large Y. Mambrini 
{\small and} C. Mu\~noz
}
\vspace{1.5cm}\\
Departamento de F\'isica Te\'orica C--XI and Instituto de F\'isica
Te\'orica C--XVI,
\\
Universidad Aut\'onoma de Madrid, Cantoblanco,
28049 Madrid, Spain.
\vspace{1cm}\\
 
\end{center}

\abstract{
We analyze the indirect detection of neutralino dark matter in
supergravity scenarios with non-universal soft scalar and gaugino
masses.
In particular, the gamma-ray flux arising from the galactic center
due to neutralino annihilation is computed.
In important regions of the parameter space it can be 
increased significantly (about two orders of magnitude) 
with respect to the universal scenario.
This result is compared with the sensitivity of current and
planned experiments, such as the satellite-based detectors 
EGRET and GLAST and the ground-based atmospheric telescope
HESS.
For example, for $\tan\beta< 35$ the sensitivity region of GLAST and HESS
can only be reached in non-universal scenarios.



}

\vspace{3cm}

\newpage

\tableofcontents








\newpage

\setcounter{page}{1}
\pagestyle{plain}

\section{Introduction}

There are substantial evidences suggesting the existence
of dark matter in the Universe.
In addition to the observations  
of galactic rotation curves \cite{Persic}, cluster of
galaxies,
and large scale flows \cite{Freedman},
implying
$0.1\lsim \relic h^2\lsim 0.3$,
the recent data obtained by the WMAP satellite \cite{wmap03-1}
confirm that dark matter must be present with
$0.094\lsim\relic h^2\lsim 0.129$.

The lightest neutralino, 
$\tilde{\chi}^0_1$,
a particle predicted by the
supersymmetric (SUSY) extension of the standard model,
is one of the leading candidates for dark matter.
It is specially interesting because, in addition to be usually
stable, it is a 
weakly interacting massive particle (WIMP),
and therefore it
can be present in the
right amount to explain the observed matter density.

There are promising methods for the
indirect detection of WIMPs through the analysis of their annihilation
products \cite{Feng}. One of them consists of detecting the gamma rays produced by
these annihilations in the galactic halo \cite{nuevas}-\cite{Mambrini}.
For this one uses atmospheric Cherenkov telescopes or space-based
gamma-ray detectors. 
In fact, 
one of the space-based experiments,
the Energetic Gamma-Ray Experiment Telescope (EGRET) 
on the Compton Gamma-Ray Observatory 
detected a signal \cite{EGRET} that apparently cannot be explained with the 
usual gamma-ray background. In particular, after
five years of mapping the gamma-ray sky up to an energy
of about 20 GeV, there are evidences of a source of 
gamma rays above about 1 GeV in the galactic center region
with a value for the flux of about
$10^{-8}$ cm$^{-2}$\ s$^{-1}$.
It is worth noticing, however, that alternative explanations
for this result have been proposed 
modifying conveniently  
the standard theory of galactic gamma ray \cite{modi}.

Projected experiments might
clarify the situation. For example,
a larger sensitivity will be provided by the Gamma-ray Large Area
Space Telescope (GLAST) \cite{GLAST}, which is scheduled for launch
in 2006. 
In addition, energies
as high as 300 GeV can be measured testing models with heavy 
dark matter more
easily than EGRET.
Even if alternative explanations 
solve the data problem pointed out by EGRET, 
GLAST will be able to detect a smaller flux of gamma rays from
dark matter $\sim10^{-10}$ cm$^{-2}$\ s$^{-1}$.
Besides,
ground-based atmospheric Cherenkov telescopes
are complementary to space-based detectors.
In particular, the High Energy Stereoscopic System (HESS) 
experiment \cite{HESS}, which has begun operations with four
telescopes, 
has a large effective area of $~10^{5}$  m$^2$ (versus 1 m$^2$ for GLAST) 
allowing the detection of very high energy fluxes $\gsim 1$ TeV.
On the contrary, the energy threshold is larger, $E_{\gamma}> 60$ GeV.


Given the experimental situation described above, concerning the
detection of dark matter, and assuming that this 
is made of neutralinos, it is natural to wonder how big 
the annihilation cross section for its indirect 
detection can be.
Obviously, this analysis is crucial in order to know the 
possibility of detecting dark matter in the experiments.

This has been carried out \cite{Feng,Ullio,Boer,Wang,farrill}
in the usual minimal supergravity (mSUGRA) scenario, where the soft terms
of the minimal supersymmetric standard model (MSSM) 
are assumed to be universal at the unification scale, 
$M_{GUT} \approx 2\times  10^{16}$ GeV, 
and radiative electroweak symmetry
breaking is imposed.
Neutralino annihilations in the galactic center region 
seem not to be able
to explain the excess detected by EGRET. 
In fact, 
only
small regions of the
parameter space of mSUGRA may be compatible with the sensitivity of
GLAST
and HESS, for standard dark matter density profiles.
Clearly, in this case, 
more sensitive detectors
producing further data are needed.
However, this result might be modified by taking into account
possible departures from the mSUGRA scenario.
In particular, a more general situation in the context
of SUGRA than universality, the presence of non-universal
soft scalar and gaugino masses, might increase the annihilation 
cross section significantly, producing larger gamma-ray fluxes. 

The aim of this paper
is to investigate this general case, where non-universalities are
present in the scalar and gaugino sectors\footnote{For
analyses in the context of the effMSSM scenario,
where the parameters are defined directly at the electroweak scale,
see e.g. the recent works \cite{hooper,Bottino}.}, and
to carry out a detailed analysis of 
the prospects
for the indirect detection of neutralino dark matter in these
scenarios.
In the light of the recent experimental results,
we will be specially interested in studying how big the gamma-ray 
flux from the galactic center
can be.
Our purpose is to provide a general analysis which can be used in the
study of any concrete model.

In this sense, it is worth noticing that
such a non-universal structure 
can be recovered in the low-energy limit of some
phenomenologically appealing string
scenarios.
This is the case, 
for example, 
of heterotic orbifold models \cite{StringReview}
or D-brane constructions in the type I 
string \cite{Rigolin}.
An analysis of the gamma-ray fluxes
in these scenarios might be very interesting \cite{us}.
On the other hand, AMSB scenarios in an effective heterotic framework
were studied in Ref.~\cite{Mambrini, Mambrini2}.

The paper is organized as follows.
In Section~2 we will discuss in general the
gamma-ray detection from dark matter annihilation, 
paying special attention to its halo model dependence.
The particle physics dependence will be reviewed in
Section~3 for the case of neutralinos in mSUGRA.
In Section~4 we will study the general case where 
scalar and gaugino non-universalities are present. 
In particular, we will indicate
the conditions under which a significant enhancement of the resulting
gamma-ray flux is obtained. 
Finally, the conclusions are left for Section~5.

\section{Gamma-ray flux from dark matter annihilation}



Annihilation of dark matter particles 
in the galactic 
center \cite{center}, in the galactic
halo \cite{halo2}
or in the haloes of nearby galaxies \cite{nearby,nearby2} 
could produce detectable fluxes of high-energy photons.
We are interested in the analysis of fluxes from the galactic center
because they can be significantly enhanced for some specific
dark matter density profiles \cite{Ullio3}, as we will see below.

There are two possible types of gamma rays that can be produced
by the annihilation. First, gamma-ray lines
from processes $\chi \chi \to \gamma \gamma$ 
\cite{Ullio1} and $\chi \chi \to \gamma Z$ \cite{Ullio2}.
This signal would be very clear since the photons are
basically mono-energetic. Unfortunately,
the neutralino does not couple directly to the photon,
the Feynman diagrams are loop suppressed, and therefore
the flux would be small \cite{Berezinsky}.
For more recent analyses of this possibility see Refs.~\cite{Ullio3,Wang}.
On the other hand, continuum gamma rays produced
by the decay of neutral pions generated in the cascading
of annihilation products
will give rise to
larger fluxes \cite{Salati}.
Although 
we will concentrate on the latter possibility in the following, let us
remark
that we have also checked for all cases studied in the paper
the flux of monochromatic gamma rays.
This turns out to be too small for a standard NFW density profile, 
with an upper bound of about
$10^{-13}$ cm$^{-2}$\ s$^{-1}$, and therefore much below the sensitivity of GLAST.

For the continuum of gamma rays, the
differential flux 
coming from a direction forming an angle
$\psi$ with respect to the galactic center is
\begin{equation}
\frac{d \Phi_{\gamma}}{d \Omega d E}=\sum_i \frac{1}{2}\frac{dN_{\gamma}^i}{dE_{\gamma}}
 \langle\sigma_i v\rangle \frac{1}{4 \pi m_{\chi}^2}\int_{line\ of\ sight} \rho^2
\ dl\ ,
\label{Eq:flux}
\end{equation}
\noindent 
where the discrete sum is over all dark matter annihilation
channels,
$dN_{\gamma}^i/dE_{\gamma}$ is the differential gamma-ray yield,
$\langle\sigma_i v\rangle$ is the annihilation cross section averaged over its
velocity
distribution, $m_{\chi}$ is the mass of the dark matter particle,
and $\rho$ is the dark matter density. 
It is worth noticing here that 
we have included in the above equation the
factor 1/2 omitted in previous literature\footnote{As discussed in
Ref.~\cite{nearby2} the simplest way to see this is 
realizing
that $\sigma$ is the annihilation cross section for a given pair of
particles.
Thus in a given volume with $N$ WIMPs the annihilation rate will be
$\sigma v$ times the number of pairs $N(N-1)/2$.
In the continuum limit this result can be approximated as $\sigma
v n^2/2$.}.
Assuming a spherical halo,
$\rho=\rho(r)$ with the galactocentric distance
$r^2=l^2+R_0^2-2lR_0 \cos \psi$,
where $R_0$ is the solar distance to the galactic center
($\simeq$ 8 kpc \cite{radiosolar}).
Following Ref.~\cite{Ullio3}, one can separate in the above equation
the particle physics part from the halo model dependence introducing the
(dimensionless) quantity
%
\begin{equation}
J(\psi)=\frac{1}{8.5 ~\mathrm{kpc}}
\left(
\frac{1}{0.3 ~\mathrm{GeV/cm^3}}
\right)^2
\int_{line\ of\ sight}\rho^2(r(l,\psi))\ dl\ .
\end{equation}
%

\noindent Thus the gamma-ray flux can be expressed as 
\begin{eqnarray}
\Phi_{\gamma}(E_{thr})
& = & 
1.87\times 10^{-13}\ \mathrm{cm^{-2}\ s^{-1}} 
\nonumber \\
&\times & \mbox{}
\frac{1}{2}\sum_i
\int_{E_{thr}}^{m_{\chi}}dE_{\gamma}\frac{dN_{\gamma}^i}{dE_{\gamma}}
\left(
\frac{\langle\sigma_i 
v\rangle}{10^{-29} {\mathrm{cm^3 s^{-1}}}}
\right)
\left(
\frac{100 ~\mathrm{GeV}}{m_{\chi}}
\right)^2
\bar{J}(\Delta \Omega) \Delta \Omega\ .
\label{Eq:totflux}
\end{eqnarray}

\noindent
Here $E_{thr}$ is the lower threshold energy of the detector,
and with respect to the upper limit of the integral notice that
neutralinos move at galactic velocity and therefore
their annihilation occurs at rest. The quantity 
$\bar{J}(\Delta \Omega)$ is defined as
\begin{equation}
\bar{J}(\Delta \Omega)\equiv \frac{1}{\Delta \Omega}\int _{\Delta \Omega}
J(\psi)\ d\Omega\ .
\end{equation}
\noindent Although the flux is maximized in the direction of the
galactic center (corresponding to $\psi=0$), rather than
$J(0)$ one must consider the integral of $J(\psi)$
over the spherical region of solid angle 
$\Delta \Omega$ given by the angular acceptance of the detector
which is pointing towards the galactic center.
Typically $\Delta \Omega$ is about $10^{-3}$ sr.


\subsection{Astrophysics contribution}




A crucial ingredient for the calculation of 
$\bar{J}$, and therefore of the flux of gamma rays,
is the dark matter density profile of our galaxy. 
The different profiles that have been proposed in the literature
can be parameterised as \cite{model}
\begin{equation}
\rho(r)= \frac{\rho_0  [1+(R_0/a)^{\alpha}]^{(\beta-\gamma)/\alpha}  }{(r/R_0)^{\gamma} 
[1+(r/a)^{\alpha}]^{(\beta-\gamma)/\alpha}}\ ,
\label{profile} 
\end{equation}
where $\rho_0$ is the local (solar neighborhood) 
halo density 
and 
$a$ is a characteristic length.
Although we will use $\rho_0=0.3$ GeV/cm$^3$ throughout the paper,
since this is just a scaling factor in the analysis,
modifications to its value can be straightforwardly taken into account
in the results.
Obviously, for $r=R_0$ one recovers the local density in
Eq.~(\ref{profile}).

For ($\alpha, \beta, \gamma$)=($2,2,0$) one obtains the simple
Isothermal halo model 
$\rho(r)= \rho_0 \frac{a^2+ R_0^{2}}{a^2+r^{2}}$, where $a$ is the
core radius.
This profile is clearly non-singular at the galactic center $r=0$.
However, highly cusped profiles seems to be deduced
from N-body simulations
\footnote{For analytical derivations see e.g. the recent work \cite{Hansen}, and references therein.}.
Navarro, Frenk and White (NFW) \cite{Navarro:1996he} 
obtained ($\alpha, \beta, \gamma$)=($1,3,1$),
producing a profile with a behaviour $1/r$ at small
distances.
A more singular behaviour,  $1/r^{1.5}$, was obtained by
Moore et al. \cite{Moore:1999gc} with
($\alpha, \beta, \gamma$)=($1.5,3,1.5$).
On the other hand,
Kravtsov et al. \cite{Kravtsov:1997dp} obtained a
mild singularity towards the galactic center since for them
($\alpha, \beta, \gamma$)=($2,3,0.4$).
Moreover, Navarro et al. \cite{navarron} have recently proposed
a non-singular distribution
\begin{equation}
\rho(r)=
\rho_{-2}\ \mathrm{exp}\left\{-\frac{2}{\alpha}\left[\left(\frac{r}{r_{-2}}\right)^{\alpha}-1\right]
\right\}\ ,
\label{navarronu} 
\end{equation}
with typical values 
$\alpha=0.142$, $r_{-2}=26.4$ kpc, $\rho_{-2}=0.035$ GeV cm$^{-3}$.

Clearly, the situation concerning the halo models is still unclear.
The situation is similar from the observational viewpoint.
Reconstructions of the observed rotation curves 
are claimed to be consistent or inconsistent with
the predictions of N-body simulations 
depending on the authors.

%
\begin{table}
\begin{tabular}{|c|cccc|}
\hline 
&a (kpc)&$\overline{J}\left( 10^{-2}\right)$& $\overline{J}\left(
  10^{-3}\right)$ &
$\overline{J}\left( 10^{-5}\right)$ \\
\hline 
Isothermal & 3.5& 2.439 $\times 10^1$ & 2.466  $\times 10^1$  & 2.470  $\times 10^1$ \\
Kravtsov et al.& 10 & 1.625  $\times 10^1$  & 1.932 $\times 10^1$   & 2.368  $\times 10^1$ \\
NFW& 20& 3.665  $\times 10^2$  &  1.223  $\times 10^3$ & 1.261 $\times 10^4$  \\
Navarro et al.&  & 0.952 $\times 10^3$  & 3.015 $\times 10^3$   & 1.407  $\times 10^4$ \\
Moore et al.& 28 & 2.069 $\times 10^4$   & 1.443 $\times 10^5$  & $1.033 \times 10^7$ \\
\hline 
\end{tabular}
\centering
\caption{Different dark matter density profiles with
the corresponding value of $\bar{J}(\Delta \Omega)$, for $\Delta
\Omega=10^{-2},10^{-3},10^{-5}$ sr.}
\label{tab}
\end{table}
%

Taking all the above into account, we show in Table~\ref{tab}
the value of 
$\bar{J}(\Delta \Omega)$, with $\Delta \Omega=10^{-2},10^{-3},10^{-5}$ sr,
for the most widely
used dark matter profiles.
In the computation one usually assumes a constant density for
$r<0.01$ pc, 
in order to avoid
the problem of the divergence of the singular profiles.
 Obviously, the gamma-ray flux can be largely enhanced 
depending on the profile chosen\footnote{
We are conservative and do not consider here 
the effect on this study
of the possible existence
of a super massive black hole at the center of our galaxy
of $3.7\times 10^6$
solar masses \cite{schodel}. 
This might produce a `spike' in the dark matter density
profile leading to behaviour $1/r^{2.4}$ \cite{Gondolo:1999ef}, and therefore 
increasing substantially the gamma-ray flux \cite{Bertone:2002je}.}.
For example, for $\Delta \Omega=10^{-3}$ sr, 
the flux produced by
the Moore et al. profile 
is about two orders of magnitude larger than the one produced
by the NFW, and this is about two orders of magnitude larger
than the flux produced by the Isothermal and Kravtsov profiles.

For the sake of definiteness, we will use in our analysis
a NFW  
profile
with $\Delta \Omega=10^{-3}$ sr.
Using Table~\ref{tab}, 
the results can easily be extended to other profiles
and angular resolutions, since
$\bar J(\Delta \Omega)$ enters just as a scaling factor in the gamma-ray flux
(\ref{Eq:totflux}). 


\begin{figure}
\begin{center}
\centerline{
\epsfig{file=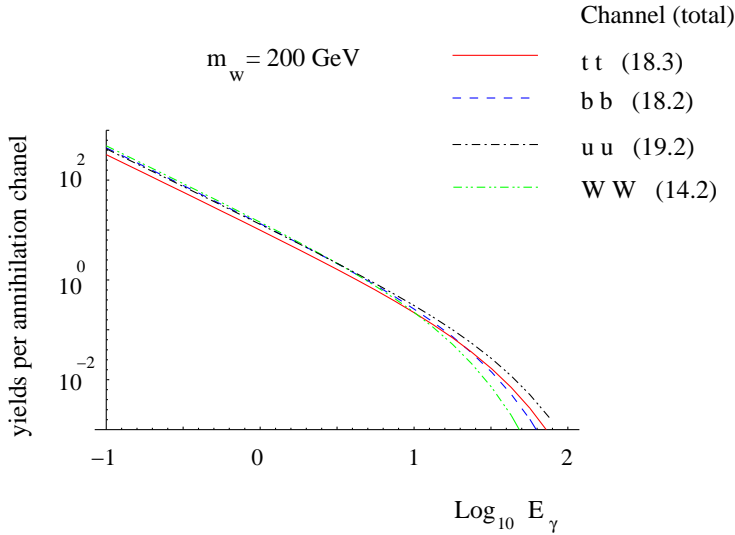,width=0.6\textwidth}\hskip 1cm}
\caption{
{\footnotesize {Differential photon yield 
$dN_{\gamma}/dE_{\gamma}$ 
for several
annihilation channels, using a WIMP with mass $m_{w}=200$ GeV.
The total photon yield for each channel is also shown in parenthesis.}}}
\label{fig:gammadistribution}
\end{center}
\end{figure}

\subsection{Particle physics contribution}

Concerning the
particle physics contribution to the gamma--ray flux
(\ref{Eq:totflux}), one has all the other factors in front of 
$\bar{J}(\Delta \Omega)$.
Their values will depend on the
specific theoretical scenario where the WIMPs are
embedded. 
With respect to the differential photon yield, it is known
that this shows scaling properties \cite{Salati}, i.e. it depends only on the scaled
variable
$x=E_{\gamma}/m_{w}$, where $m_{w}$ is the mass of the WIMP. Thus 
can be reasonably parameterised for each
possible annihilation channel of the WIMPs  \cite{Salati,Ullio3}.
For example, assuming that WIMPs 
annihilate producing
$WW$, $ZZ$, $b\overline{b}$, $t\overline{t}$, and $u\overline{u}$,
one can write
$dN_{\gamma}/dx=a e^{-bx}/x^{1.5}$,
with
$(a,b)=(0.73,7.76)$ for the annihilation channels
$WW$ and $ZZ$ \cite{Ullio3,nullio}, $(1.0,10.7)$ for $b\overline{b}$,
$(1.1,15.1)$ for $t\overline{t}$, and $(0.95,6.5)$ for $u\overline{u}$
\cite{Feng}.
In fact,
the spectrum of gamma-rays is very
close for all the above channels as illustrated in 
Fig.~\ref{fig:gammadistribution}, 
except in the case $\tau ^+ \tau ^-$ which produces smaller values 
(see e.g. \cite{Salati,Ullio}). 
In our computation of the gamma-ray flux below
we use 
the more exact PYTHIA fragmentation 
Monte Carlo \cite{pythia}.


On the other hand, the influence of the theoretical framework
is crucial
for the value of the annihilation cross section $\sigma_i$.
As mentioned in the Introduction, we will work in the context
of the simplest SUSY construction, 
where the lightest 
supersymmetric particle (LSP) 
is absolutely stable, and therefore a candidate for dark matter.
It is remarkable that in most of the parameter
space of this construction the LSP is an electrically neutral 
(also with no strong interactions) particle, called neutralino.
This is welcome since otherwise the LSP 
would bind to nuclei and would be excluded as a candidate
for dark matter from unsuccessful searches for exotic heavy 
isotopes.

As a matter of fact, 
there are four neutralinos, $\tilde{\chi}^0_i~(i=1,2,3,4)$, 
since they are the physical 
superpositions of the fermionic partners of the neutral electroweak 
gauge bosons, 
called bino ($\tilde{B}^0$) and wino ($\tilde{W}_3^0$), and of the
fermionic partners of the  
neutral Higgs bosons, called Higgsinos ($\tilde{H}^0_u$, 
$\tilde{H}_d^0$). 
Therefore the lightest neutralino, $\tilde{\chi}^0_1$, will be the 
dark matter candidate.
In the basis ($\tilde B^0$, $\tilde W_3^0$, $\tilde H_u^0$,
$\tilde H_d^0$),
the neutralino mass matrix is given by
%
\begin{equation}
\arraycolsep=0.01in
{\cal M}_N=\left( \begin{array}{cccc}
M_1 & 0 & -m_Z\cos \beta \sin \theta_W^{} & m_Z\sin \beta \sin \theta_W^{}
\\
0 & M_2 & m_Z\cos \beta \cos \theta_W^{} & -m_Z\sin \beta \cos \theta_W^{}
\\
-m_Z\cos \beta \sin \theta_W^{} & m_Z\cos \beta \cos \theta_W^{} & 0 & -\mu
\\
m_Z\sin \beta \sin \theta_W^{} & -m_Z\sin \beta \cos \theta_W^{} & -\mu & 0
\end{array} \right)\;,
\label{eq:matchi}
\end{equation}
%
where $M_1$ and $M_2$ are the bino and wino masses respectively, 
$\mu$ is the Higgsino mass parameter 
and $\tan\beta= \langle H_u^0\rangle/\langle H_d^0\rangle$ 
is the ratio of Higgs vacuum expectation values. 
This can be diagonalized by a matrix $Z$ such that we can express
the lightest neutralino as
\begin{equation}
\tilde{\chi}^0_1 = {Z_{11}} \tilde{B}^0 + {Z_{12}} \tilde{W}_3^0 +
{Z_{13}} \tilde{H}^0_d + {Z_{14}} \tilde{H}^0_u\ .
\label{lneu}
\end{equation}
It is commonly defined that $\tilde{\chi}^0_1$  is mostly gaugino-like 
if $P\equiv \vert {Z_{11}} \vert^2 + \vert {Z_{12}}  \vert^2 > 0.9$, 
Higgsino-like
if $P<0.1$, and mixed otherwise.

\begin{figure}
\centerline{
\epsfig{file=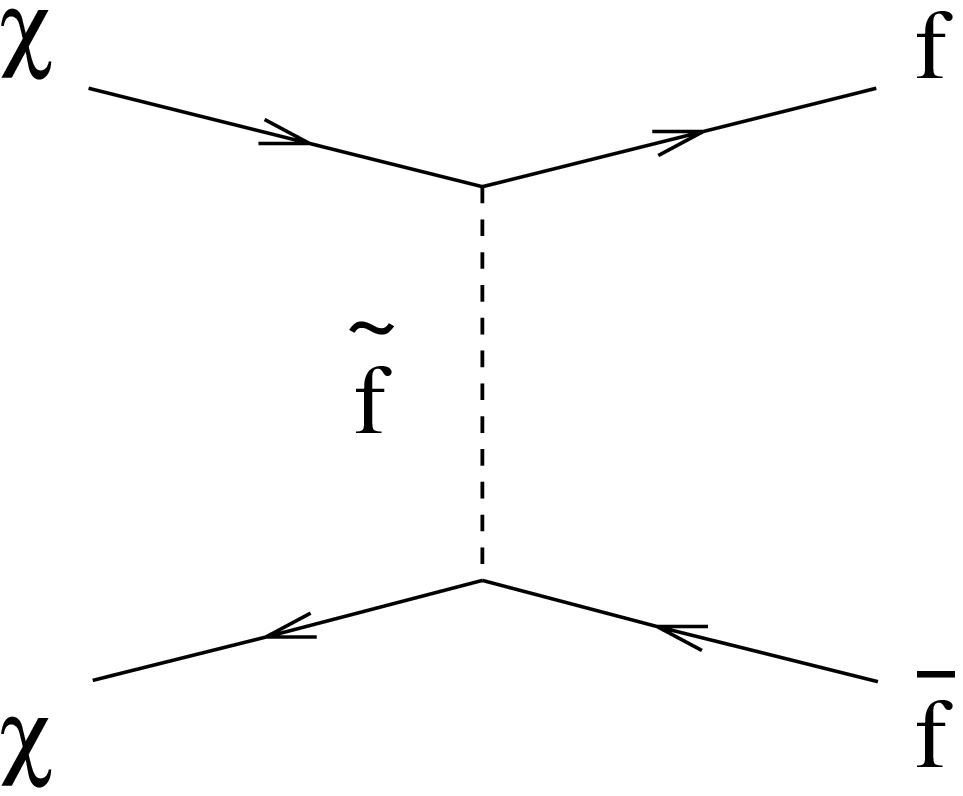,width=0.15\textwidth}\hskip 1cm
    \hspace{1cm}     \epsfig{file=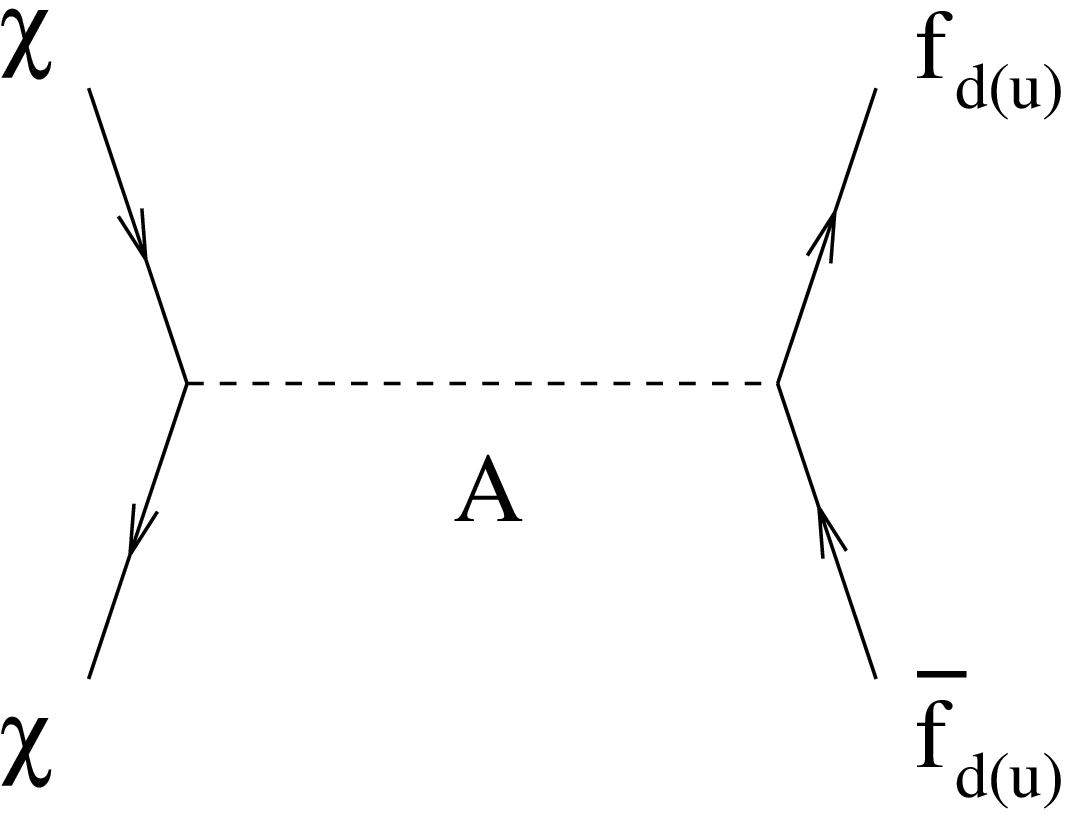,width=0.15\textwidth}\hskip 1cm
       \epsfig{file=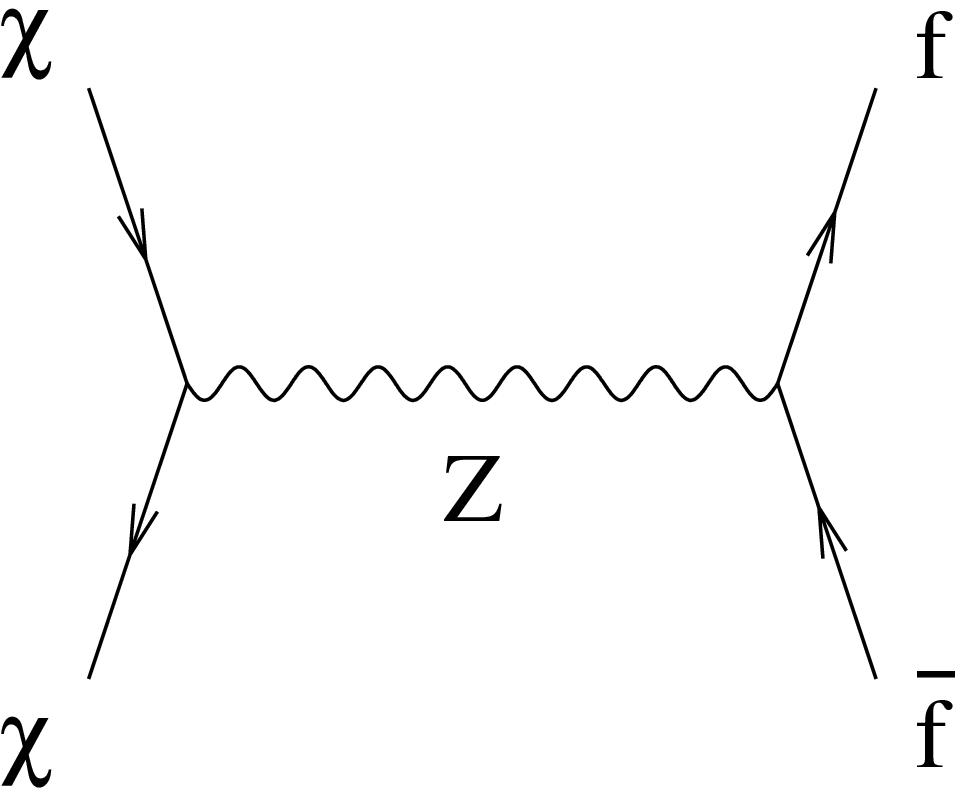,width=0.15\textwidth}\hskip 1cm
      \hspace{1.5cm} 
\epsfig{file=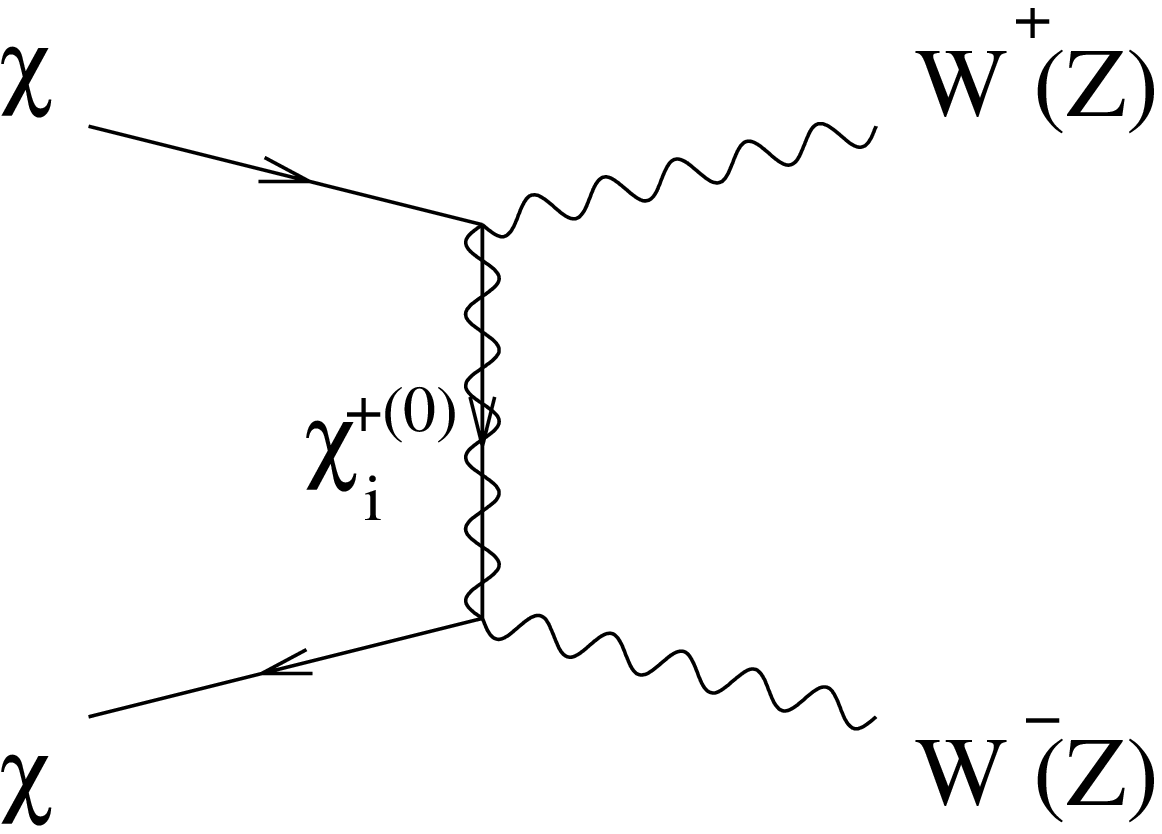,width=0.15\textwidth}\hskip 1cm
        }

 $\propto \frac{m_{\chi} m_f}{m_{\tilde f}^2}Z_{11}^2~~~$
$\propto \frac{m_{\chi}^2}{m_A^2}
\frac{Z_{11}Z_{13,14}}{m_W}  m_{f_d} \tan \beta (\frac{m_{f_u} }{\tan\beta})~ 
~~$
$\propto \frac{m_f m_{\chi}}{m_Z^2} Z_{13,14}^2~$
$\propto \frac{[-Z_{14} V_{21}^* + \sqrt{2} Z_{12} V_{11}^*]^2 (-Z_{13} N_{31}^* +  
Z_{14} N_{41}^*)^2}{1+m_{\chi_i^{+(0)}}^2 /
m_{\chi}^2 - m_{W(Z)}^2 / m_{\chi}^2}$
          \caption{{\footnotesize Dominant neutralino annihilation
diagrams. Relevant parts of the amplitudes are shown explicitly
Terms between parenthesis correspond to $f_u$ and $Z$ final states
in second and fourth diagrams. V and N are chargino and neutralino
mixing
matrices.
}}
        \label{fig:feynmandetail}

\end{figure}

In Fig.~\ref{fig:feynmandetail} we show the relevant 
Feynman diagrams contributing to neutralino annihilation.
As we will see, the cross section
can be significantly enhanced depending on the
SUSY model under consideration.
We will concentrate here on the 
SUGRA scenario, where the soft terms are
determined at the unification 
scale, $M_{GUT}\approx 2\times10^{16}$ GeV, after SUSY breaking,
and radiative electroweak symmetry breaking is imposed.

Let us start reviewing the situation in the case of 
mSUGRA, where the soft terms are assumed to be universal.

\section{mSUGRA predictions  for the gamma-ray flux
}



Let us first recall that in mSUGRA
one has only four free parameters defined at the GUT scale:
the soft scalar mass $m$, the soft gaugino mass $M$, 
the soft trilinear coupling $A$, and 
$\tan\beta$.
In addition, the sign of the Higgsino mass parameter, $\mu$,
remains also undetermined by the 
minimization of the Higgs potential, which at tree level implies
\begin{equation}
  \mu^2 = \frac{\higgsdew - \higgsuew \tan^2 \beta}{\tan^2 \beta -1 } - 
  \frac{1}{2} M_Z^2\ .
  \label{electroweak}
\end{equation}

As is well known, in this scenario the lightest neutralino
is mainly bino \cite{binor,seealso}. 
To understand this result qualitatively recall the evolution
of $m_{H_u}^2$ towards generically large
and negative values with the scale.
Since 
$\mu^2$ given by Eq.~(\ref{electroweak}),
for reasonable values of  $\tan\beta$,
can be approximated as,
\begin{equation}
\mu^2\approx -m_{H_u}^2-\frac{1}{2} M_Z^2\ ,
\label{electroweak2}
\end{equation}
then it becomes also 
large.
In particular, $|\mu|$ becomes 
much larger than
$M_1$ and $M_2$. Thus, as can be easily understood from 
Eqs.~(\ref{eq:matchi}) and (\ref{lneu}),
the lightest neutralino will be mainly gaugino, and in particular
bino, since
at low energy $M_1=\frac{5}{3}\tan^2\theta_W M_2\approx 0.5 M_2$.

Now, using 
Eq.~(\ref{Eq:totflux}) one can compute the
gamma-ray flux for different values of the 
parameters. 
As a consequence of the $\tilde{\chi}^0_1$  being mainly bino,
the predicted $\sigma_i$ is small, and therefore the flux is below
the present accessible experimental regions.
Note that only  ${Z_{11}}$ is large and then the contribution of 
diagrams in Fig.~\ref{fig:feynmandetail} will be generically small.
In addition,
the (tree-level)
mass of the CP-odd Higgs, $A$,
\begin{equation}
m^2_A=m_{H_d}^2+m_{H_u}^2+2\mu^2\ ,
\label{ma} 
\end{equation}
will be large. 
The previous formula can be rewritten as
\begin{equation}
  m^2_A\approx m_{H_d}^2-m_{H_u}^2-M_Z^2\ ,
  \label{ma2} 
\end{equation}
where  Eq.~(\ref{electroweak2}) has been used.
As mentioned above,
$m_{H_u}^2$ evolves at low energy towards large and negative values,
producing a large value for $m_A$, and therefore 
a further suppression in the annihilation channels.

\begin{figure}
    \begin{center}
\centerline{
       \epsfig{file=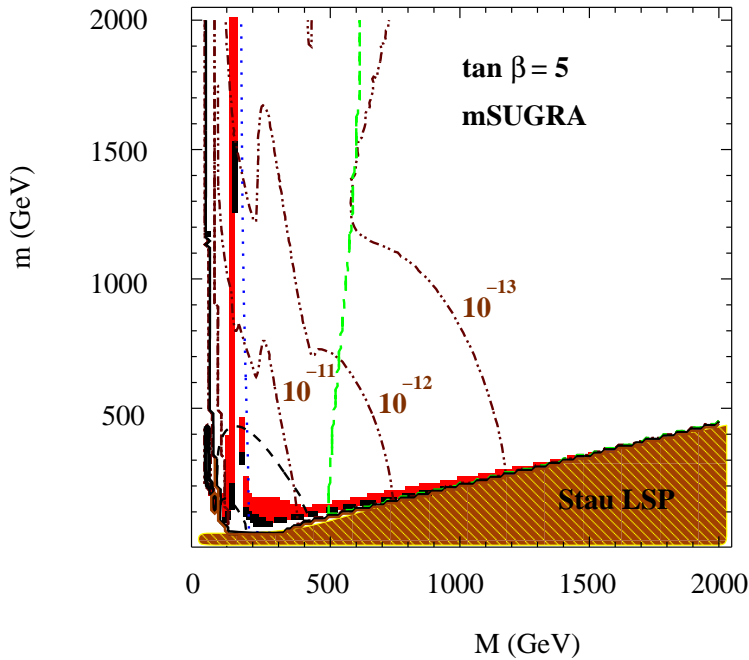,width=0.6\textwidth}
       }
          \caption{{\footnotesize 
Gamma-ray flux $\Phi_{\gamma}$
for a threshold of 1 GeV,
in the parameter space of the mSUGRA scenario ($m$, $M$) 
for $\tan \beta=5$, $A=0$
and $\mu > 0$.
The 
dot-dashed curves are contours of $\Phi_{\gamma}$
in cm$^{-2}$ s$^{-1}$ units.
A NFW profile is used with $\Delta \Omega=10^{-3}$ sr.
The region to the left of the 
light grey (green) dashed line
is excluded by the lower bound 
on the Higgs mass.
The region to the left of the dotted line
is excluded by the lower bound on the chargino mass
$m_{\tilde\chi_1^{\pm}}>103$ GeV.
The region to the right of the black dashed line
corresponds to
$\asusy< 7.1\times10^{-10}$, and would be excluded by
$e^+e^-$ data.
The 
region at the bottom (Stau LSP) is excluded because the lightest stau
is the LSP.
The dark grey (red) region fulfils
$0.1\leq \Omega_{\tilde{\chi}_1^0}h^2\leq 0.3$ (the black region on top of this
indicates the WMAP range $0.094<\Omega_{\tilde{\chi}_1^0}h^2<0.129$).
}}
        \label{fig:univtb5scan}
    \end{center}
\end{figure}

This fact is shown in 
Fig.~\ref{fig:univtb5scan}
using the DarkSUSY program \cite{darksusy}\footnote{
As discussed in 
Eq.~(\ref{Eq:flux}), we are including  
in the computation
the factor 1/2 which was missing in previous literature,
and also in the equations
of the program. 
We have used the Fortran code SuSpect2 \cite{Suspect} to solve the renormalization 
group equations (RGE) for the soft supersymmetry breaking terms between the GUT scale and the 
electroweak scale. These parameters are then passed on to the {\bf C}
code 
micrOMEGAs \cite{micromegas}
to perform the calculation of physical masses for the superpartners and various indirect
 constraints to be described below.
We have compared our results with the ones obtained with the recent
released version 
of DarkSUSY \cite{darksusynew}, and they are very similar. Only for large
values of $\tan\beta$ and $M$ there is a factor $\sim3$  of discrepancy in the flux.
}.
There, contours 
of the gamma-ray flux  $\Phi_{\gamma}$ in the
parameter space ($m$, $M$) for $\tan \beta=5$, $A=0$
and $\mu > 0$
are plotted,
using the complete one-loop corrected
effective potential. 
We will not consider in the calculation the opposite sign of $\mu$
because this would produce a negative contribution for the $g_{\mu}-2$,
and, as will be discussed below, we are mainly interested in positive
contributions. Recall that the sign of the contribution is basically
given by $\mu M_2$, and that $M$, and therefore $M_2$, can always
be made positive after performing an $U(1)_R$ rotation.
As input for the top mass, $m_t(pole)=175$ GeV
has been used.

In the
computation we are using a NFW profile 
with $\Delta \Omega=10^{-3}$ sr, and recent experimental and astrophysical
constraints are taken into account. In particular, the lower
bounds on the masses of the supersymmetric particles and on the
lightest Higgs have been implemented, as well as the experimental
bounds on the branching ratio of the \bsg\ process and on 
$\asusy$. Due to its relevance, the effect of the WMAP constraint on
the dark matter relic density is shown explicitly. 

Concerning $\asusy$, we have taken into account the
recent experimental result for the muon
anomalous magnetic moment \cite{g-2}, as well as the most recent
theoretical evaluations of the Standard Model contributions
\cite{newg2}. It is found that when $e^+e^-$ data
are used the experimental excess in $(g_\mu-2)$ would constrain a
possible supersymmetric contribution to be
$\asusy=(27.1\pm10)\times10^{-10}$. 
At $2\sigma$ level this implies 
$7.1\times10^{-10}\lsim\asusy\lsim47.1\times10^{-10}$.
It is worth noticing here that when tau data are used a smaller
discrepancy with the experimental measurement is found.
In order not to exclude the latter possibility we will discuss the 
relevant value
$\asusy= 7.1\times10^{-10}$ in the figures throughout the paper. 
The region of the parameter space to the right of this line
will be forbidden (allowed) if one consider electron (tau) data.

On the other hand,
the measurements of $B\to X_s\gamma$ decays 
at 
CLEO \cite{cleo} 
and BELLE \cite{belle},
lead to bounds on the branching ratio 
$b\to s\gamma$. In particular we impose on our computation
$2.33\times 10^{-4}\leq BR(b\to s\gamma)\leq 4.15\times10^{-4}$, where the evaluation 
is carried out using the routine provided by
the program micrOMEGAs \cite{micromegas}.
This program is also used for our evaluation of $\asusy$ and 
relic neutralino density. 

As we can see in Fig.~\ref{fig:univtb5scan}, 
the points corresponding to a sufficiently low relic neutralino density,
and satisfying the Higgs
mass bound\footnote{Let us remark that this bound must be relaxed
accordingly when departures from mSUGRA are considered.} in mSUGRA, 
$m_h>114$ GeV,
are located 
in the narrow coannihilation branch of the parameter space, 
i.e. the region
where the stau is the next to the LSP producing efficient coannihilations.
For these points the neutralino mass is larger than the top mass
(note that we can deduce its value in the
plot from the value of $M$, since in mSUGRA 
$m_{\tilde{\chi}^0_1}\simeq {M_1}\simeq 0.4\ M$
implying $m_{\tilde{\chi}^0_1}\gsim 200$ GeV),
and $t \overline{t}$ final states are allowed.
In fact, 
through the $A$-exchange channel,
these are the main source of the gamma-ray flux.
The contribution from $b\overline{b}$ states is also relevant,
and the relation between both contributions can be obtained taking into account
that (see Fig.~\ref{fig:feynmandetail})
$\sigma_A(t\overline{t})\propto \beta_t m_t^2/\tan^2\beta$,
$\sigma_A(b\overline{b})\propto \beta_b m_b^2\tan^2\beta$, with
$\beta_f=\sqrt{1-m_f^2/m_{\chi}^2}$,
and therefore
$\sigma_A(t\overline{t})/
\sigma_A(b\overline{b})\sim
\frac{1}{\tan^4\beta}(m_t/m_b)^2$
for sufficiently large neutralino mass.
In particular, for $\tan\beta=5$ one obtains
$\sigma_A(t\overline{t})/
\sigma_A(b\overline{b})\sim 2$.
There is of course the possibility of   
$\tau \overline{\tau}$ final states, but their contributions are
negligible compared with the $b\overline{b}$ ones.
Note that, following the above arguments,
$\sigma_A(\tau\overline{\tau})/
\sigma_A(b\overline{b})\sim
\frac{1}{3}(m_{\tau}/m_b)^2\sim 0.05$, where 3 is a colour factor.


Let us finally remark that if we impose consistency
with the $e^+e^-$ data concerning $(g_\mu-2)$,
the whole parameter space would be excluded
(notice that the region to the right
of the black dashed line in Fig.~\ref{fig:univtb5scan} corresponds
to $\asusy<7.1\times10^{-10}$).

\begin{figure}
    \begin{center}
\centerline{
       \epsfig{file=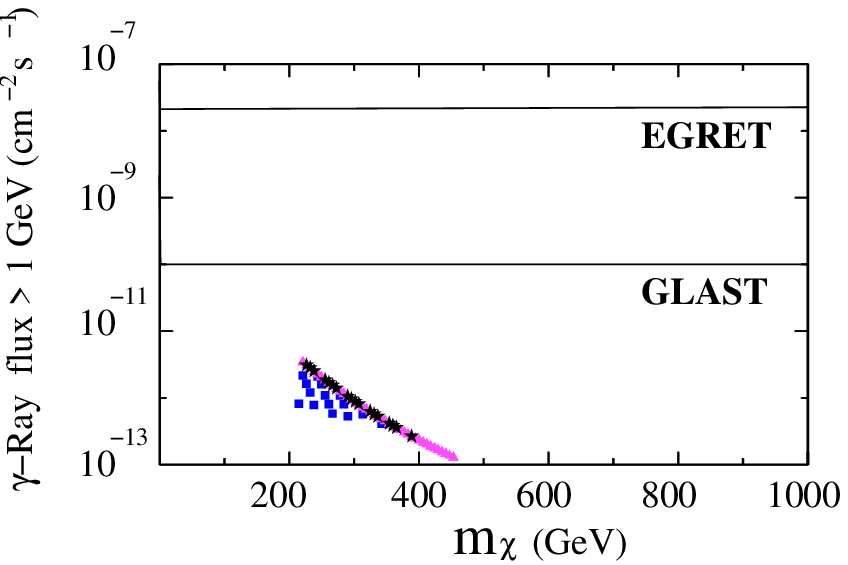,width=0.55\textwidth}\hskip 1cm
       \epsfig{file=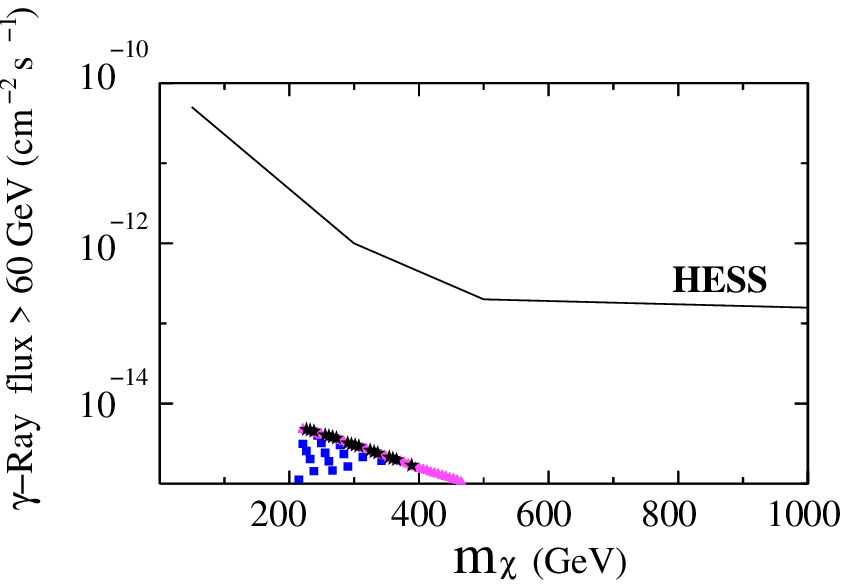,width=0.55\textwidth}
       }
          \caption{{\footnotesize
Scatter plot of the gamma-ray flux $\Phi_{\gamma}$ 
for a threshold of 1 GeV (left), and 60 GeV (right),
as a function of the neutralino mass
$m_{\chi}$,
using
the same parameter space as in Fig.~\ref{fig:univtb5scan},
for $\tan \beta=5$, $A=0$
and $\mu > 0$. 
All points have $\asusy< 7.1\times10^{-10}$, and therefore
would be excluded by ($g_{\mu}-2$) if we
impose consistency with the $e^+e^-$ data.
They fulfil
all other experimental constraints, and those 
depicted with light grey (magenta) triangles have 
$0.129<\Omega_{\tilde{\chi}^0_1}h^2<0.3$, 
with black stars have
$0.094<\Omega_{\tilde{\chi}^0_1} h^2<0.129$,
and finally with dark grey (blue) boxes have
$0.03<\Omega_{\tilde{\chi}^0_1} h^2<0.094$ 
with the appropriate rescaling of the density of neutralinos
in the galaxy as discussed in the text. 
From top to bottom, 
the first solid line in the left frame 
corresponds to the signal reported by EGRET, and
the upper area bounded by the second solid line will be analyzed by 
GLAST experiment (HESS experiment in the right frame).
}}
        \label{fig:univtb5flux}
    \end{center}
\end{figure}

We summarize the above results in Fig.~\ref{fig:univtb5flux}, using
the same parameter space as in Fig.~\ref{fig:univtb5scan}.
There, the values of $\Phi_{\gamma}$ allowed by all experimental constraints
as a function of the neutralino mass 
$m_{\tilde{\chi}^0_1}$ are depicted.
Points shown with black stars
have  $0.094<\Omega_{\tilde{\chi}^0_1} h^2<0.129$, and are therefore favoured
by WMAP.
Points with $0.129<\Omega_{\tilde{\chi}^0_1} h^2<0.3$ are shown with
light grey (magenta)
triangles.
We also consider the possibility that not all the dark matter is made
of
neutralinos, allowing
$0.03<\Omega_{\tilde{\chi}^0_1} h^2<0.094$. In this case we have 
to rescale the density of 
neutralinos in the galaxy $\rho(r)$ in Eqs.~(\ref{profile}) and
(\ref{navarronu}) by a factor
$\Omega_{\tilde{\chi}^0_1}h^2/ 0.094$. 
Points 
corresponding to this possibility are
shown with dark grey (blue) boxes.

We observe that, generically, 
the gamma-ray flux for a threshold of 1 GeV 
(left frame) is constrained
to be 
$\Phi_{\gamma} \lsim 10^{-11}$ cm$^{-2}$ s$^{-1}$.
Obviously, in this mSUGRA scenario with
$\tan\beta=$ 5 the EGRET data cannot be 
reproduced, and more sensitive detectors producing further data are needed.
Note that GLAST experiment, where values of the gamma-ray flux as low as
$10^{-10}$ cm$^{-2}$ s$^{-1}$ will be accessible, is not sufficient.
For atmospheric telescopes, such as HESS, the situation is similar,
as can be seen in the plot on the right frame.

\begin{figure}
\vskip -3cm \hskip 15cm

    \begin{center}
\centerline{
       \epsfig{file=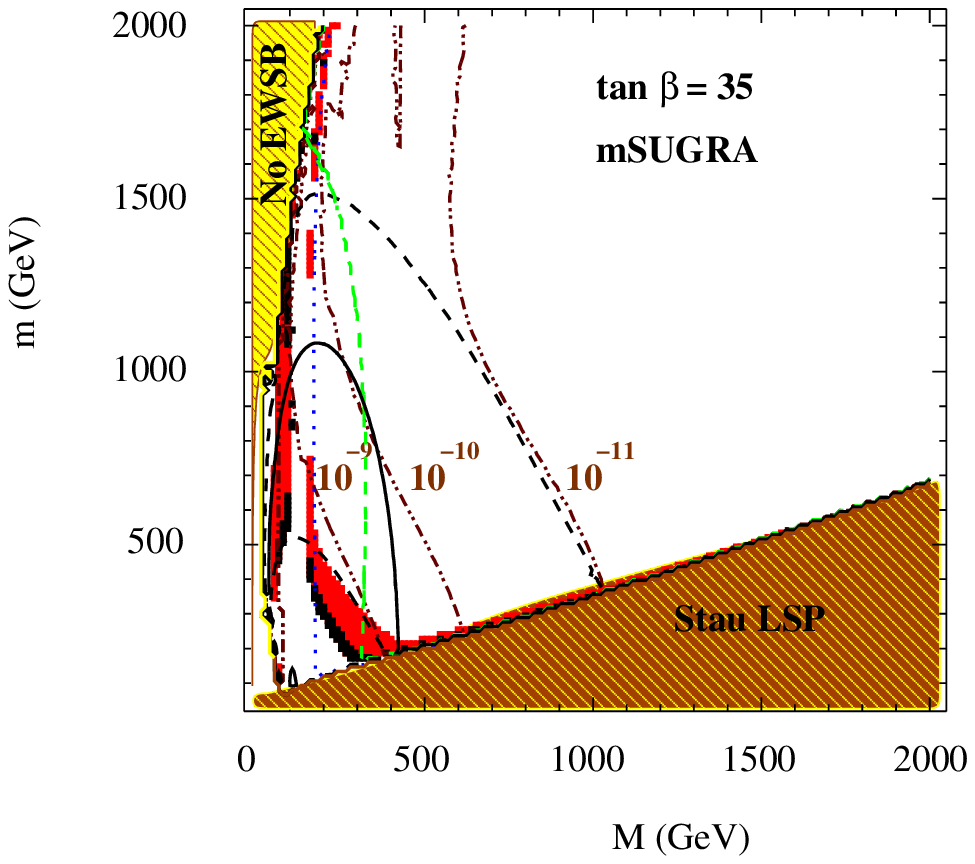,width=0.6\textwidth}
       }
          \caption{{\footnotesize
The same as in Fig.~\ref{fig:univtb5scan} but for tan$\beta$=35.
The region to the left of the solid line is excluded by $b\to s\gamma$.
}}
        \label{fig:univtb35scan}
    \end{center}
    \begin{center}
\centerline{
       \epsfig{file=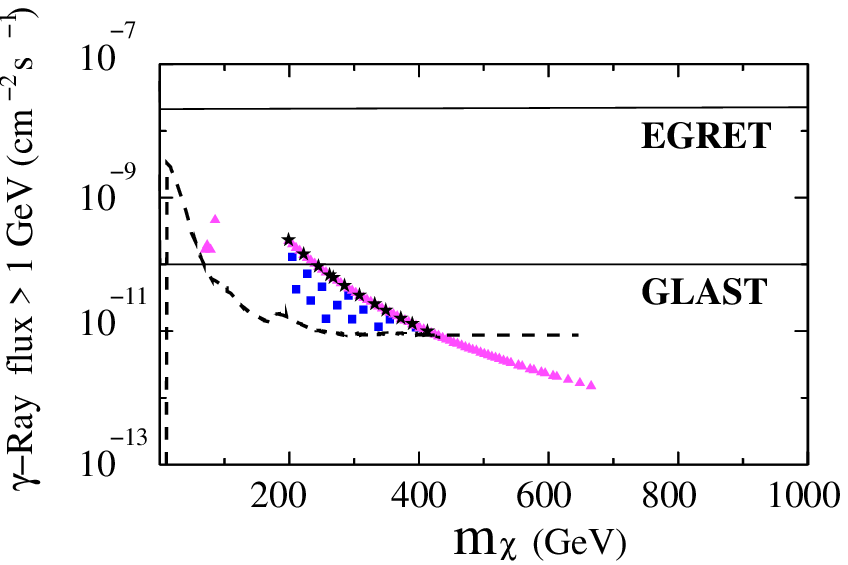,width=0.55\textwidth}\hskip 1cm
       \epsfig{file=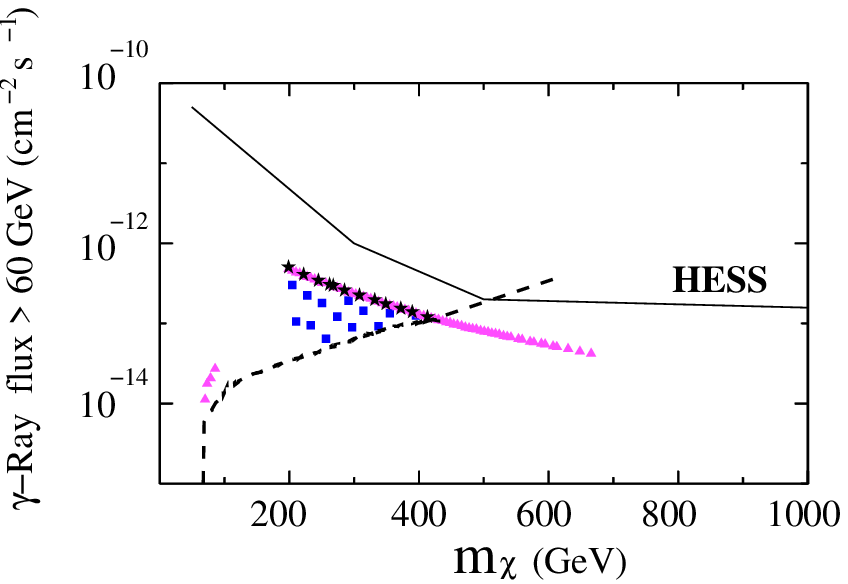,width=0.55\textwidth}
       }
          \caption{{\footnotesize The same as in 
Fig.~\ref{fig:univtb5flux} but for tan$\beta$=35. Now, only the lower
areas bounded by the dashed line correspond to 
$\asusy< 7.1\times10^{-10}$, and therefore
would be excluded by ($g_{\mu}-2$) if we
impose consistency with the $e^+e^-$ data.}}
        \label{fig:univtb35flux}
    \end{center}
\end{figure}

The annihilation cross section can be increased, 
and therefore the gamma-ray flux, 
when the value of $\tan\beta$ is increased.
Notice for instance that the contribution of the
$b\overline{b}$ final state, which is proportional
to $\tan^2\beta$ is now more important
than the $t\overline{t}$ one. For example, for  $\tan\beta=35$,
using the above approximation one obtains
$\sigma_A(t\overline{t})/
\sigma_A(b\overline{b})\approx
\frac{1}{35^4}(m_t/m_b)^2\sim 10^{-3}$.
In addition, the bottom Yukawa coupling increases, and as a consequence 
$m_{H_d}^2$ decreases, implying that
$m^2_A$, given by Eq.~(\ref{ma2}),
also decreases.
Indeed, annihilation channels through $A$ exchange are more important
now
and their contributions to the flux will increase it.
This effect is shown in Fig.~\ref{fig:univtb35scan}, which can be
compared
with Fig.~\ref{fig:univtb5scan}.

The above results concerning the gamma-ray flux are
summarized in Fig.~\ref{fig:univtb35flux}.
There we see that the flux for a threshold of 1 GeV is 
larger than for 
$\tan\beta=$5.
In any case, only a very small region of the parameter space turns out
to be accessible for GLAST,
and no region for HESS.
It is worth noticing that now only 
the lower areas bounded by the dashed line 
correspond to $\asusy< 7.1\times10^{-10}$, and therefore
would be excluded by ($g_{\mu}-2$) if we
impose consistency with the $e^+e^-$ data.

We have also analyzed very large values of $\tan\beta$,
such as 50. In this case, still EGRET data cannot be reproduced,
but larger regions of the parameter space would be accessible
for GLAST and also HESS.
In particular, fluxes with
$\Phi_{\gamma} \lsim 10^{-9}$ cm$^{-2}$ s$^{-1}$ ($\Phi_{\gamma} \lsim 10^{-11}$
cm$^{-2}$ s$^{-1}$) 
can be obtained for a
threshold of 1 GeV (60 GeV).



Let us finally remark that the result obtained above concerning EGRET,
should be expected given the following argument.
The relic density can be approximated in the simplest situation, 
when the annihilation cross
section is velocity independent, as 
$\Omega h^2 \sim \frac{3\times 10^{-27}\mathrm{cm^3 s^{-1}}}{<\sigma v>}$.
This implies that in order to obtain the
result $\Omega h^2 \sim 0.1$ one needs a cross section
$\sigma v \sim 3\times 10^{-26}\mathrm{cm^3s^{-1}}$.
Using this value for the annihilation cross section in the galaxy
in Eq.~(\ref{Eq:totflux}), for a NFW profile with 
$\overline{J}\left(10^{-3}\right)\approx 3\times 10^3$ and a photon
yield of about 10 as can be deduced from Fig.~\ref{fig:gammadistribution},
one obtains a gamma-ray flux 
$\Phi_{\gamma}\sim 10^{-9}$ cm$^{-2}$ s$^{-1}$,
i.e. below EGRET sensitivity.
In fact, the value of the flux is typically two or three orders of
magnitude smaller
than this estimation, since in the early Universe 
coannihilations can also contribute to $\sigma v$ as discussed above,
thus $3\times 10^{-26}\mathrm{cm^3s^{-1}}$ can be considered as an upper limit.
The results shown in Figs.~\ref{fig:univtb5flux} and \ref{fig:univtb35flux}
confirm this rough discussion.

\newpage

\section{Departures from mSUGRA}


As discussed in detail in Refs.~\cite{cggm03-1,Birkedal-Hansen:2002am,cggm03-12} 
in the context of direct detection,
the neutralino-proton scattering cross section can be increased in different
ways when the structure of mSUGRA for the soft terms is abandoned. 
Since the diagrams for neutralino annihilation through neutral
Higgs exchange are related to those by crossing symmetry, 
we can use the same arguments here.
In particular, it is possible to enhance the annihilation
channels involving exchange of the CP-odd Higgs, $A$, by reducing the
Higgs mass, and also by increasing the Higgsino components of
the lightest neutralino.
As a consequence, the gamma-ray flux will be increased.
A brief analysis based on the Higgs
mass parameters, $\higgsdew$ and $\higgsuew$, at the electroweak scale
can clearly show how these effects can be achieved.

First, a decrease in the value of the mass of $A$ can be
obtained by increasing $\higgsuew$ (i.e., making it less negative) 
and/or decreasing $\higgsdew$. 
This is easily understood from Eq.~(\ref{ma2}).

Second, through the increase in the value of
$\higgsuew$ an increase in the Higgsino components of
the lightest neutralino can also be achieved. 
Making $\higgsuew$ less negative, its positive contribution to 
$\mu^2$ in (\ref{electroweak2}) would be smaller.
Eventually $|\mu|$ will be of the order of
$M_{1}$, $M_{2}$ and 
$\tilde{\chi}_1^0$ will then be a mixed Higgsino-gaugino state.
Thus annihilation channels through Higgs exchange become more important
than in mSUGRA, where $|\mu|$ is large and 
$\tilde{\chi}_1^0$ 
is mainly
bino.
This is also the case for $Z$-, $\chi_1^\pm$-, and  $\tilde{\chi}_1^0$-exchange
channels.

As pointed out in Refs.~\cite{cggm03-1,Birkedal-Hansen:2002am,cggm03-12}, 
non-universal soft parameters can produce the above mentioned
effects. Let us first consider non-universalities in the scalar
masses.


\subsection{Non-universal scalars}

We can parameterise the non-universalities
in the Higgs sector, at the GUT scale, as
follows:
\begin{equation}
  m_{H_{d}}^2=m^{2}(1+\delta_{1})\ , \quad m_{H_{u}}^{2}=m^{2}
  (1+ \delta_{2})\ .
  \label{Higgsespara}
\end{equation}
Concerning squarks and sleptons we will assume
that the three generations have the
same mass structure:
\begin{eqnarray}
  m_{Q_{L}}^2&=&m^{2}(1+\delta_{3})\ , \quad m_{u_{R}}^{2}=m^{2}
  (1+\delta_{4})\ , 
  \nonumber\\
  m_{e_{R}}^2&=&m^{2}(1+\delta_{5})\ ,  \quad m_{d_{R}}^{2}=m^{2}
  (1+\delta_{6})\ , 
  \nonumber\\
  m_{L_{L}}^2&=&m^{2}(1+\delta_{7})\ .    
  \label{Higgsespara2}
\end{eqnarray}
Such a structure avoids potential problems
with flavour changing neutral 
currents\footnote{Another possibility would be to assume 
  that the first
  two generations have the common scalar mass $m$, and
  that non-universalities are allowed only for the third generation.
  This would not modify our analysis since, as we will see below, 
  only the third generation is relevant in our
  discussion.}.
Note also that whereas $\delta_{i} \geq -1 $, $i=3,...,7$, in order to
avoid an unbounded from below (UFB) direction breaking charge and
colour, 
$\delta_{1,2} \leq -1$
is possible as long as 
$m_1^2\,=\,m_{H_{d}}^2+\mu^2>0$ and $m_2^2\,=\,m_{H_{u}}^2+\mu^2>0$   
are fulfilled.

An increase in $\higgsu$ at the electroweak scale can be obviously
achieved by 
increasing its value at the GUT scale, i.e., with the choice
$\delta_2>0$. 
In addition, this is also produced when $m_{Q_{L}}^2$ and $m_{u_{R}}^2$
at $M_{GUT}$ decrease,
i.e. taking $\delta_{3,4} < 0$, due to 
their (negative) contribution proportional
to the top Yukawa coupling in the renormalization group equation (RGE)
of $m_{H_u}^2$. 
Similarly, a decrease in the value of $\higgsd$ at the electroweak
scale can be 
obtained by decreasing it at the GUT scale with $\delta_1<0$. Also,
this effect is produced when $m_{Q_{L}}^2$ and $m_{d_{R}}^2$ 
at $M_{GUT}$ increase, due to their (negative) contribution proportional
to the bottom Yukawa coupling in the RGE
of $m_{H_d}^2$.
Thus one can deduce that $m^2_A$ 
will be reduced 
by choosing also $\delta_{3,6} > 0$. 
In fact non-universality in the Higgs sector gives the most important
effect, and including the one in the sfermion sector the annihilation cross
section only increases slightly. Thus in what follows we will take
$\delta_{i}=0$, $i=3,...,7$.
Let us analyze then three representative cases:
\begin{eqnarray}
a)\,\, \delta_{1}&=&0\ \,\,\,\,\,\,\,\,,\,\,\,\, \delta_2\ =\ 1\ ,
\nonumber\\
b)\,\, \delta_{1}&=&-1\ \,\,\,\, ,\,\,\,\, \delta_2\ =\ 0\ ,
\nonumber\\
c)\,\, \delta_{1}&=&-1\ \,\,\,\, ,\,\,\, \delta_2\ =\ 1 
\ .
\label{3cases}
\end{eqnarray}

\begin{figure}
\vskip -3cm \hskip 15cm

    \begin{center}
\centerline{
       \epsfig{file=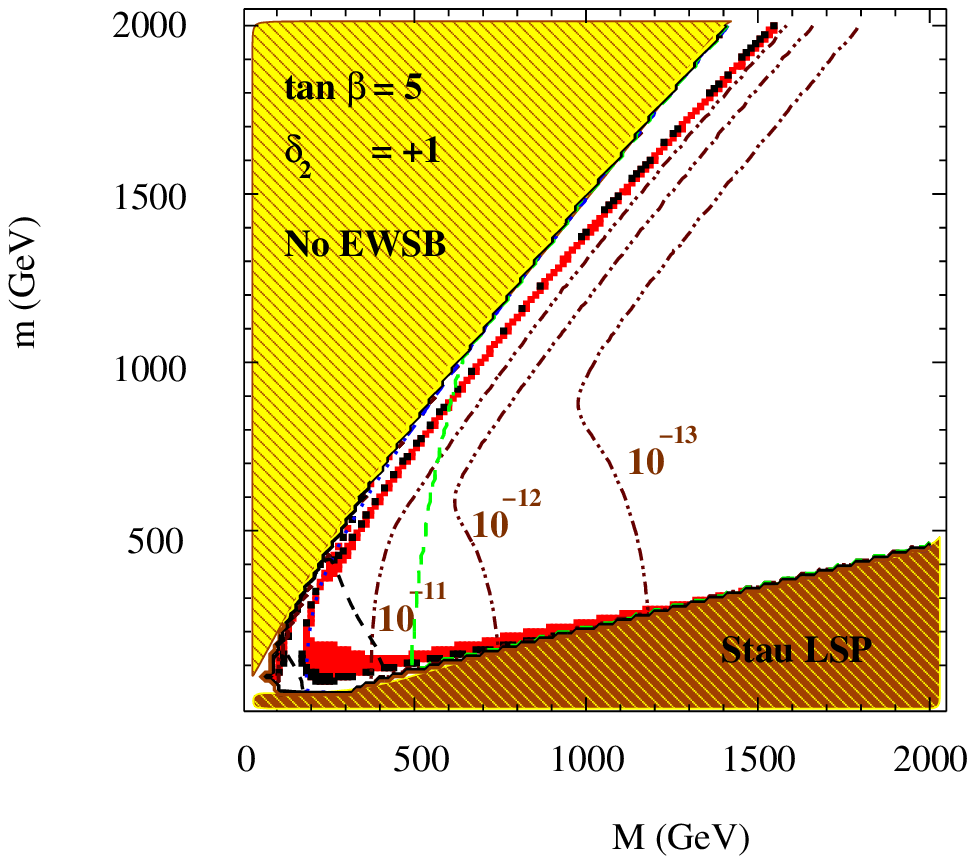,width=0.6\textwidth}
       }
          \caption{{\footnotesize The same as in 
Fig.~\ref{fig:univtb5scan} but for the non-universal scalar case {\it a)} 
$\delta_1=0$, $\delta_2=1$, discussed in Eq.~(\ref{3cases}).
}}
        \label{fig:nonuniv1tb5scan}
    \end{center}
    \begin{center}
\centerline{
       \epsfig{file=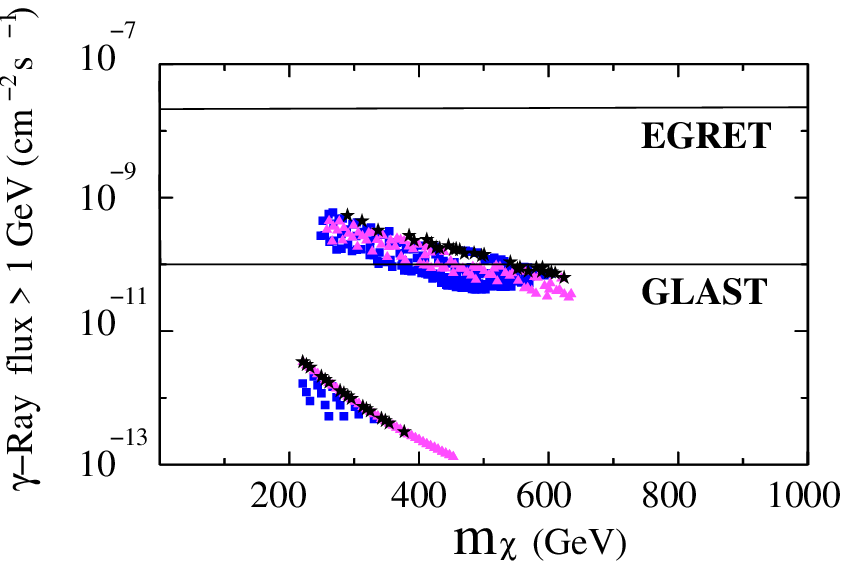,width=0.55\textwidth}\hskip 0.75cm
       \epsfig{file=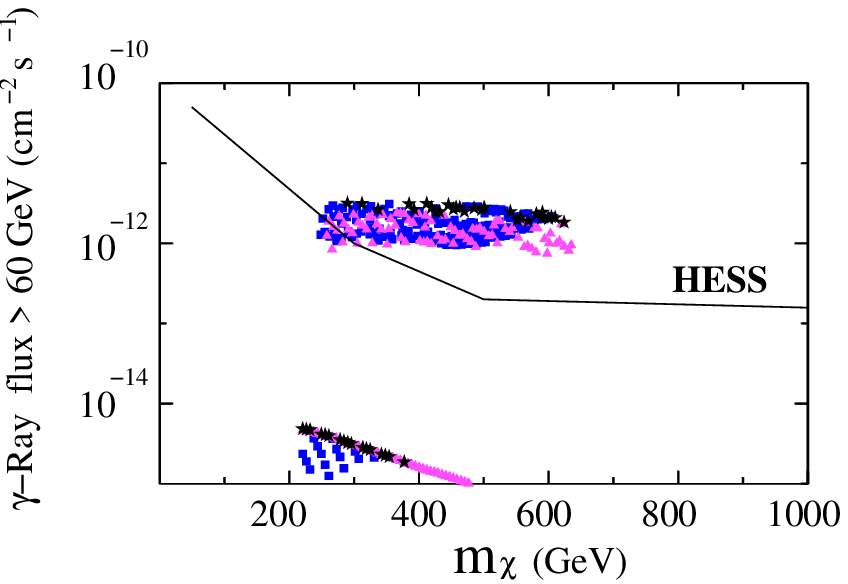,width=0.55\textwidth}
       }
          \caption{{\footnotesize 
 The same as in 
Fig.~\ref{fig:univtb5flux} but for the non-universal scalar case {\it a)} 
$\delta_1=0$, $\delta_2=1$, discussed in Eq.~(\ref{3cases}).
}}
        \label{fig:nonuniv1tb5flux}
    \end{center}
\end{figure}

Clearly, the above discussion about decreasing $\mu^2$ applies
well to case {\it a)}, where the variation in
$m_{H_u}^2$ through $\delta_2$ is relevant.
This is shown in 
Fig.~\ref{fig:nonuniv1tb5scan} for $\tan\beta=5$ and $A=0$, which
can be compared with
Fig.~\ref{fig:univtb5scan}.
Note that now 
there is an important area in the upper left (no EWSB)
where $\mu^2$ becomes negative due to the increasing in
$\delta_2$ with respect to the universal case.
We see that the decrease of the $\mu$ parameter opens
a new allowed (narrow) region of the parameter space, close to this
one.
In particular, the value of the relic density 
$\Omega_{\tilde{\chi}^0_1}$ is affected due to the increase of the
Higgsino
components of ${\tilde{\chi}^0_1}$ with respect to the dominant bino
component of the universal case. Thus the relic abundance is placed
inside the astrophysical bounds.
In Fig.~\ref{fig:feynmandetail}
we can see that this nature of the LSP 
enhances the neutralino coupling to the $Z$
(giving rise to bottom and top final states when kinematically allowed), and 
to the lightest chargino $\chi_1^\pm$ and neutralino $\tilde{\chi}_1^0$ 
(inducing $W^{\pm}W^{\pm}$ and $ZZ$ boson final states,
especially for low values of $M$).

We can see the enhancement in the gamma-ray flux 
in Fig.~\ref{fig:nonuniv1tb5flux}, where many points accessible for
GLAST and HESS are obtained.
This can be compared with the result of 
Fig.~\ref{fig:univtb5flux} for the universal scenario, where such
points are not possible
with $\tan\beta=5$.
On the other hand, the lower region, below GLAST (and HESS),
corresponds to the narrow coannihilation branch. 
In this case the neutralino is mainly bino 
and the discussion
of this region for mSUGRA in the previous section applies also here.

Let us remark, however, that if we impose consistency
with the $e^+e^-$ data concerning $(g_\mu-2)$,
the whole parameter space would be excluded
(notice that the region to the right
of the black dashed line in Fig.~\ref{fig:nonuniv1tb5scan} corresponds
to $\asusy<7.1\times10^{-10}$),
as in the case of mSUGRA.

\begin{figure}
\vskip -3cm \hskip 15cm

    \begin{center}
\centerline{
       \epsfig{file=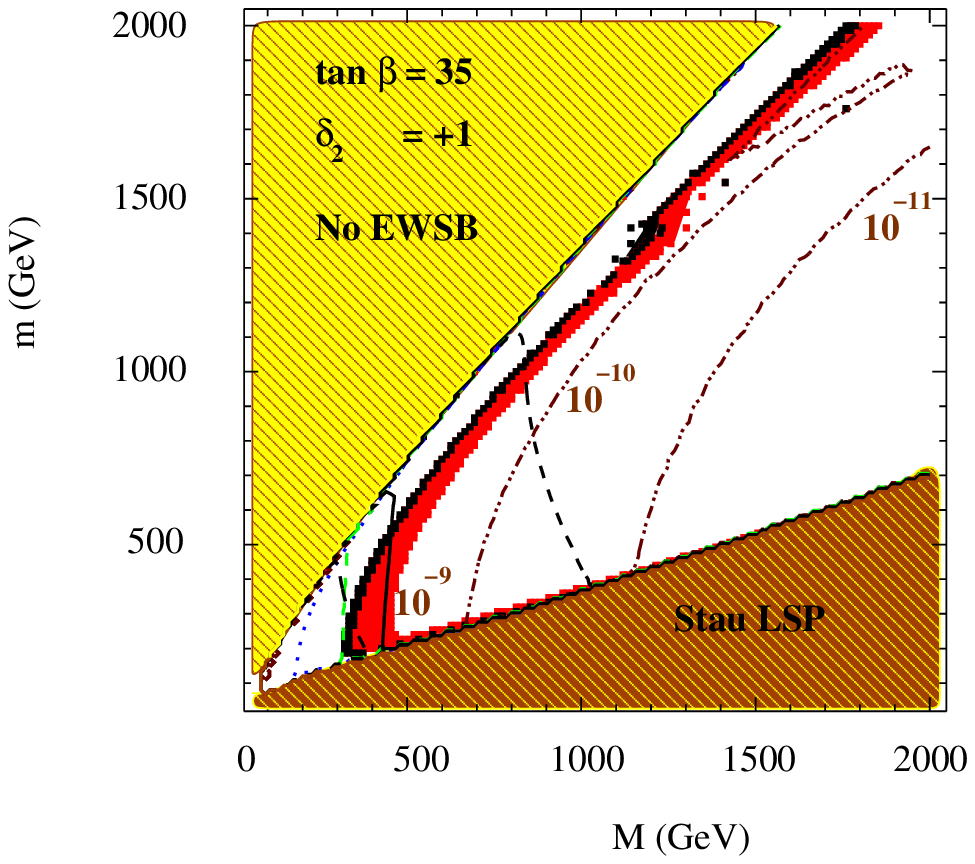,width=0.6\textwidth}
       }
          \caption{{\footnotesize 
The same as in 
Fig.~\ref{fig:univtb5scan} but for $\tan\beta=$35 and 
the non-universal scalar case {\it a)}
$\delta_1=0$, $\delta_2=1$, discussed in Eq.~(\ref{3cases}).
The region to the left of the solid line is excluded by $b\to s\gamma$.
}}
        \label{fig:nonuniv1tb35scan}
    \end{center}
\vskip -4.truecm
    \begin{center}
\centerline{
       \epsfig{file=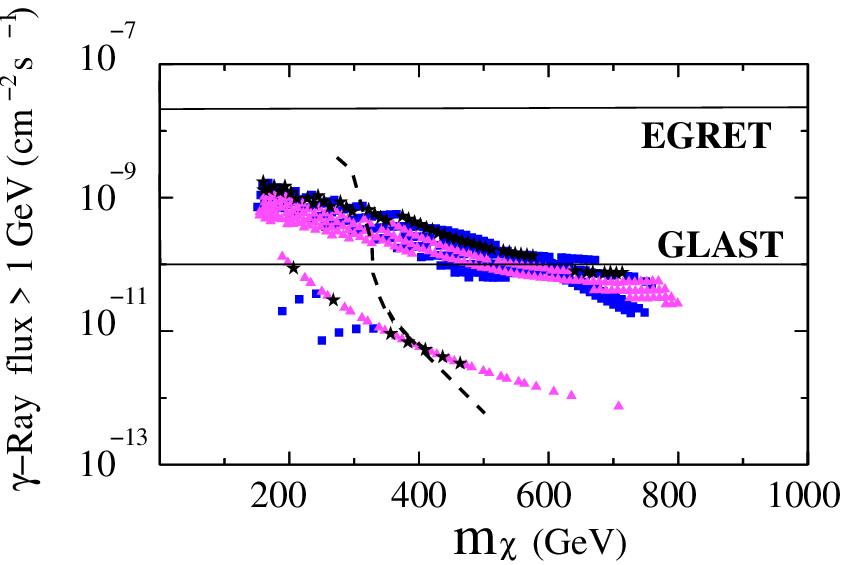,width=0.55\textwidth}\hskip 1cm
       \epsfig{file=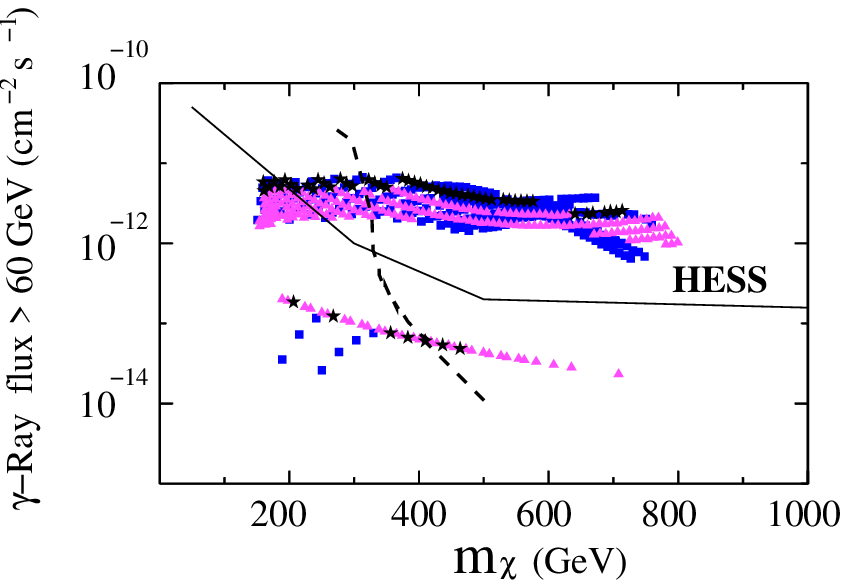,width=0.55\textwidth}
       }
          \caption{{\footnotesize 
The same as in 
Fig.~\ref{fig:univtb5flux}  but for $\tan\beta=$35 and 
the non-universal scalar case {\it a)} 
$\delta_1=0$, $\delta_2=1$, discussed in Eq.~(\ref{3cases}).
Now, the region to the right of the 
dashed line
corresponds to
$\asusy< 7.1\times10^{-10}$, and would be excluded by
$e^+e^-$ data.
}}
        \label{fig:nonuniv1tb35flux}
    \end{center}
\end{figure}

For $\tan\beta=35$, one can see in Figs.~\ref{fig:nonuniv1tb35scan}
and \ref{fig:nonuniv1tb35flux} the 
enhancement of the flux related to the coannihilation branch.
As discussed in the previous section for mSUGRA, the increase is about two 
orders of magnitude with respect to the case $\tan\beta=5$
(see Fig.~\ref{fig:nonuniv1tb5flux}), 
as expected by the couplings proportionals to tan$\beta$ 
in the $A$-exchange channels. 
On the other hand, one can also see in Fig.~\ref{fig:nonuniv1tb35flux} that 
the flux in the Higgsino branch
does not increase significantly. This is 
because it is mainly produced by the $Z$ exchange that does not depend on
tan$\beta$ at all. 
Concerning this branch, it is worth noticing that it is not so close
to the no EWSB area as in the case of mSUGRA. 
In particular, one can see in Fig.~\ref{fig:nonuniv1tb35scan}
that larger values
of the parameters imply a larger gap between the Higgsino branch
and the no EWSB area. Compare for example the points  
with $m$=1000 GeV and 2000 GeV, and with the relic density
inside the astrophysical bounds.
In order to obtain the same value of the relic density, 
for $m$=2000 GeV  a higher value
of $\mu $ is necessary.


It is worth noticing that now, unlike  the $\tan\beta=5$ case, not all points would be excluded
by ($g_{\mu}-2$) if we
impose consistency with the $e^+e^-$ data.
In Fig.~\ref{fig:nonuniv1tb35flux} those to the left of
the dashed line 
correspond to $\asusy> 7.1\times10^{-10}$, and therefore
would be allowed on this ground.

\begin{figure}
\vskip -3cm \hskip 15cm

    \begin{center}
\centerline{
       \epsfig{file=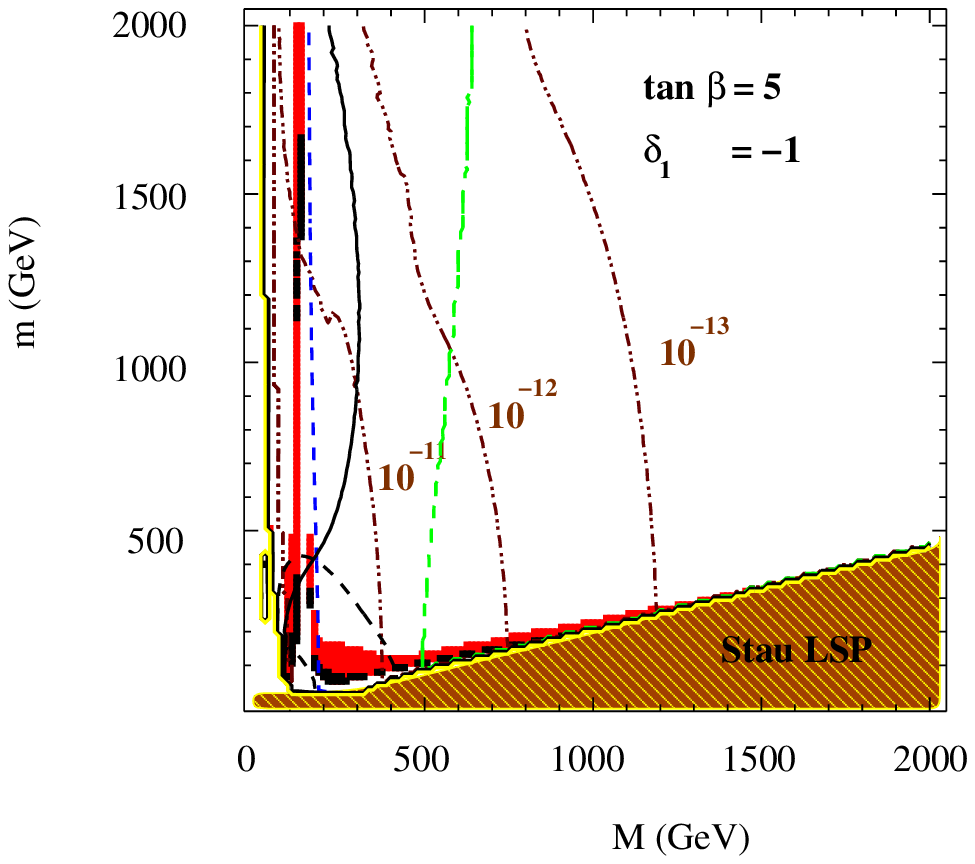,width=0.6\textwidth}
       }
          \caption{{\footnotesize 
The same as in 
Fig.~\ref{fig:univtb5scan} but for the non-universal scalar case {\it b)} 
$\delta_1=-1$, $\delta_2=0$, discussed in Eq.~(\ref{3cases}).
The region to the left of the solid line is excluded by $b\to s\gamma$.
}}
        \label{fig:nonuniv2tb5scan}
    \end{center}
    \begin{center}
\centerline{
       \epsfig{file=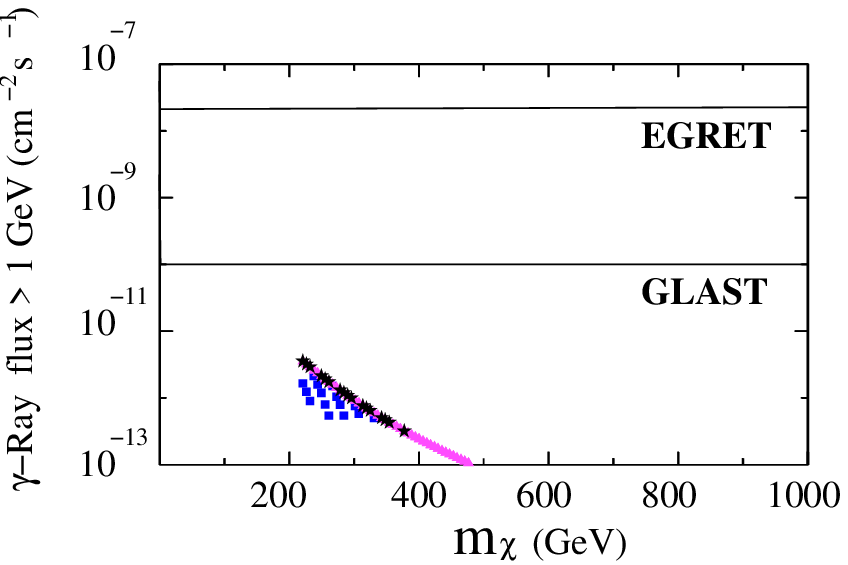,width=0.55\textwidth}\hskip 0.75cm
       \epsfig{file=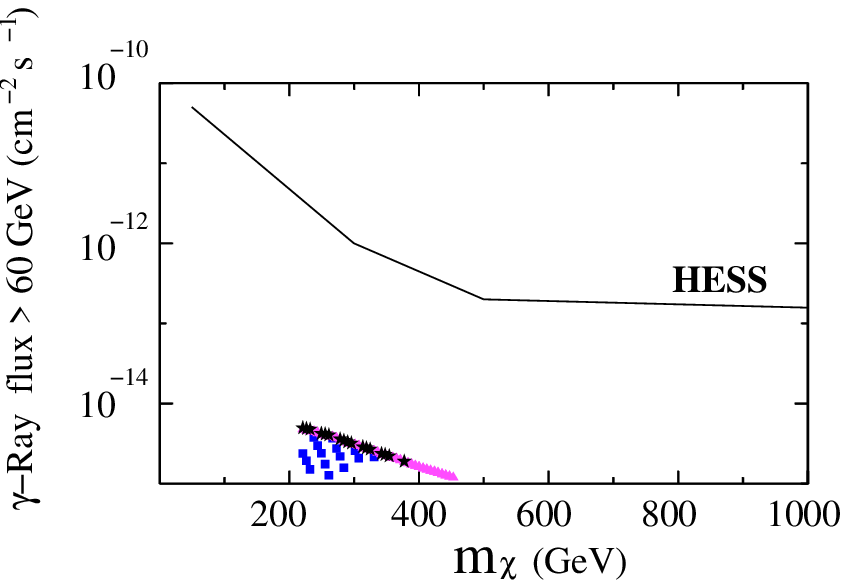,width=0.55\textwidth}
       }
          \caption{{\footnotesize 
The same as in 
Fig.~\ref{fig:univtb5flux} but for the non-universal scalar case {\it b)} 
$\delta_1=-1$, $\delta_2=0$, discussed in Eq.~(\ref{3cases}).
}}
        \label{fig:nonuniv2tb5flux}
    \end{center}
\end{figure}

\begin{figure}
\vskip -3cm \hskip 15cm

    \begin{center}
\centerline{
       \epsfig{file=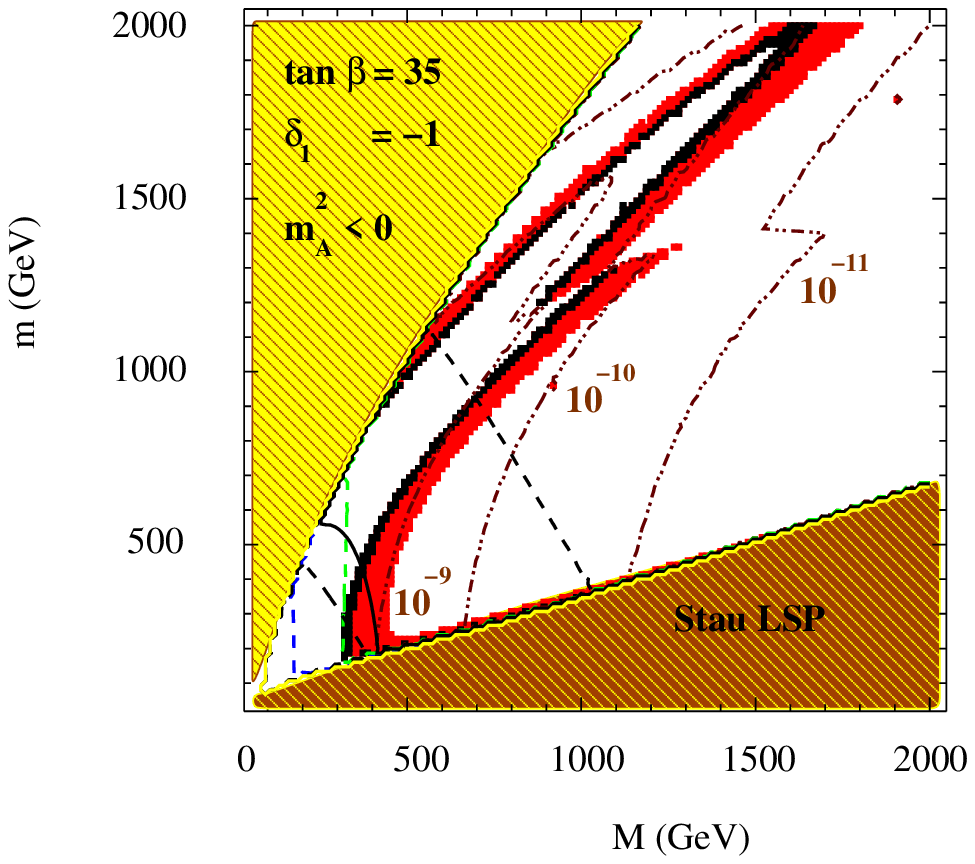,width=0.6\textwidth}
       }
          \caption{{\footnotesize 
The same as in 
Fig.~\ref{fig:univtb5scan} but for $\tan\beta=35$ and 
the non-universal scalar case {\it b)} 
$\delta_1=-1$, $\delta_2=0$, discussed in Eq.~(\ref{3cases}).
The region to the left of the solid line is excluded by $b\to s\gamma$.
 }}
        \label{fig:nonuniv2tb35scan}
    \end{center}
    \begin{center}
\centerline{
       \epsfig{file=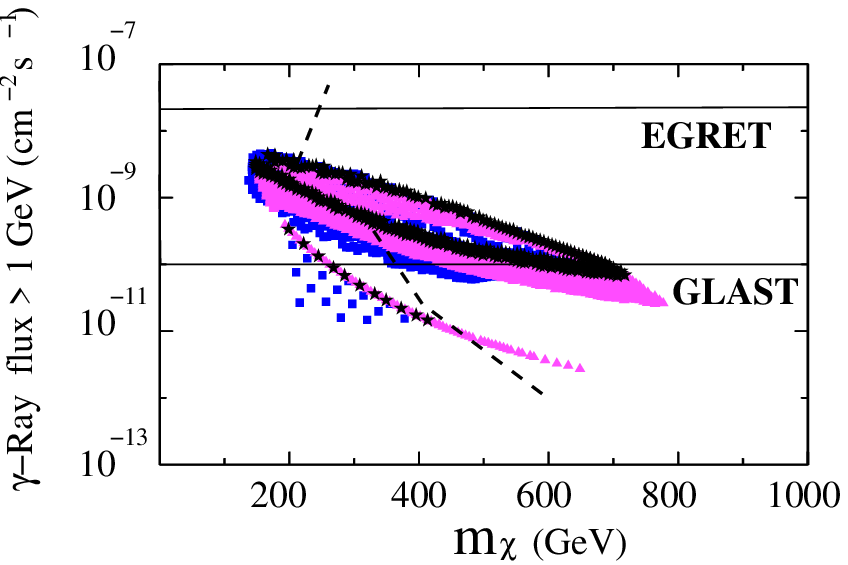,width=0.55\textwidth}\hskip 1cm
       \epsfig{file=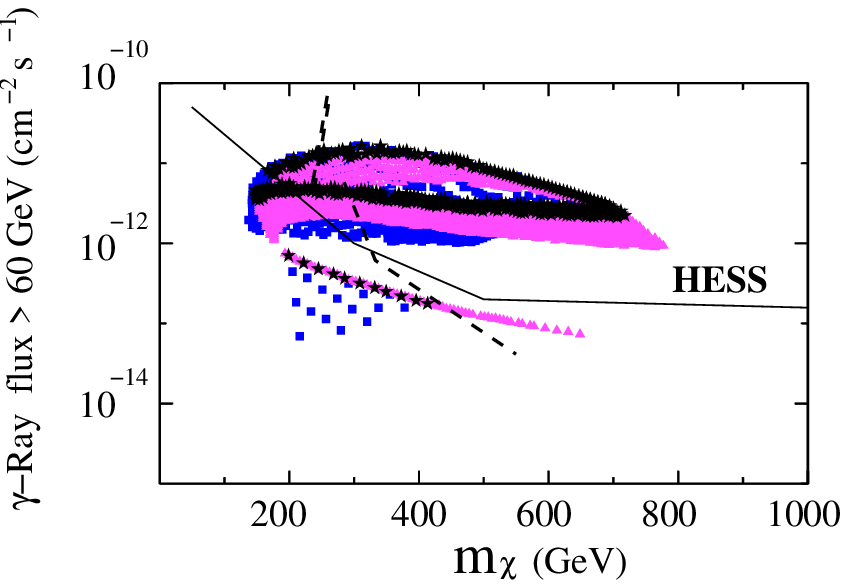,width=0.55\textwidth}
       }
          \caption{{\footnotesize 
The same as in 
Fig.~\ref{fig:univtb5flux} but for $\tan\beta=35$ and 
the non-universal scalar case {\it b)} 
$\delta_1=-1$, $\delta_2=0$, discussed in Eq.~(\ref{3cases}).
Now, the regions to the right of the dashed lines
correspond to
$\asusy< 7.1\times10^{-10}$, and would be excluded by
$e^+e^-$ data. These do not include the points 
with black stars between both dashed
lines, which are allowed.
 }}
        \label{fig:nonuniv2tb35flux}
    \end{center}
\end{figure}

On the other hand,
for
case {\it b)} with $\tan\beta=5$, studied in
Figs.~\ref{fig:nonuniv2tb5scan}
and \ref{fig:nonuniv2tb5flux},
the only region which is allowed by the experimental constraints 
is the one corresponding to the narrow coannihilation branch.
Recall that now $m_{H_u}^2$ is not varied and therefore the effect of
decreasing $\mu^2$ is not present.
As in the case of mSUGRA, the gamma-ray flux is small and below
the accessible experimental areas.
However, for $\tan\beta=35$,
due to the effect of decreasing  $m_{H_d}^2$ significantly,
$m_A^2$ can be very small. Notice 
the area in the upper left of 
Fig.~\ref{fig:nonuniv2tb35scan} where
$m_A^2$ becomes negative. 
As a consequence, Fig. \ref{fig:nonuniv2tb35flux} shows similar
fluxes to those of the case {\it a)} in
Fig. \ref{fig:nonuniv1tb35flux}, but with 
a larger 
contribution from the $A$ exchange.
Thus
interesting regions of the parameter space are accessible for
future experiments.

Now we can see in Fig.~\ref{fig:nonuniv2tb35scan} 
(and  Fig. \ref{fig:nonuniv2tb35flux})
that there are two cosmologically allowed corridors.
This can be easily understood from the variation in $m_A$.
For example, in  Fig.~\ref{fig:nonuniv2tb35scan},
for a given value of $m$, and from large to small $M$,
$m_A$ decreases. Starting with $M=2000$ GeV, one has $m_A>2m_{\tilde{\chi}^0_1}$,  and
therefore
a small annihilation cross section producing a 
large
relic density. However, 
when $m_A$ becomes close to $2m_{\tilde{\chi}^0_1}$
the annihilation cross section increases and therefore the
relic density decreases entering in the first allowed corridor with
$0.094<\Omega_{\tilde{\chi}^0_1}h^2<0.3$.
For slightly smaller values of $M$ one finally arrives to the A-pole region where
$m_A\sim 2m_{\tilde{\chi}^0_1}$ and
the relic density is too small.
Smaller values of $m_A$ close this region, producing a decrease in the
annihilation cross section, which allows to 
enter in the second corridor with the relic density inside the
observational bounds. For $m_A$ too small this second corridor is closed,
and the relic density is too large,
$\Omega_{\tilde{\chi}^0_1}h^2>0.3$.
Finally, one arrives to the region where 
$m_A^2$ becomes negative.

\begin{figure}
\vskip -3cm \hskip 15cm

    \begin{center}
\centerline{
       \epsfig{file=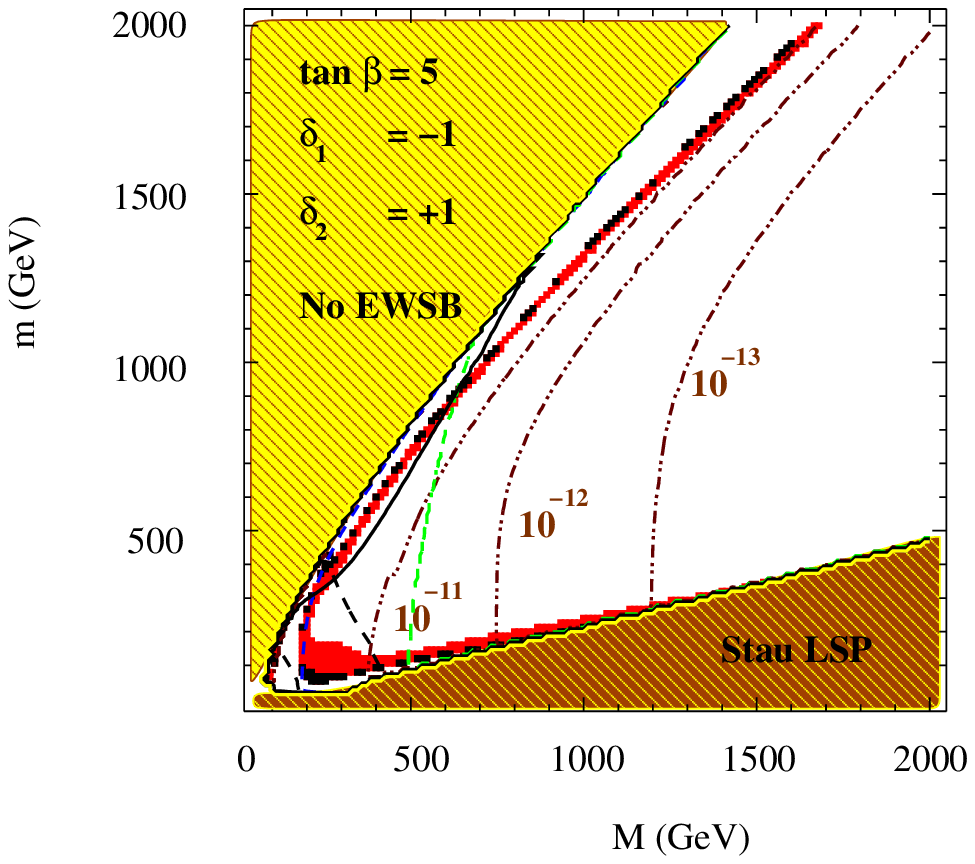,width=0.6\textwidth}
       }
          \caption{{\footnotesize 
The same as in 
Fig.~\ref{fig:univtb5scan} but for  
the non-universal scalar case {\it c)} 
$\delta_1=-1$, $\delta_2=1$, discussed in Eq.~(\ref{3cases}).
The region to the left of the solid line is excluded by $b\to s\gamma$.
}}
        \label{fig:nonuniv3tb5scan}
    \end{center}
    \begin{center}
\centerline{
       \epsfig{file=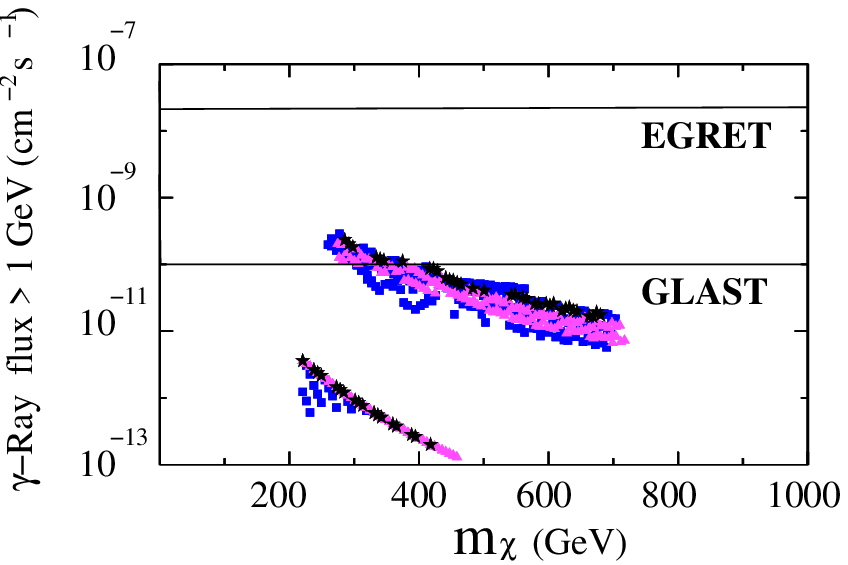,width=0.55\textwidth}\hskip 1cm
       \epsfig{file=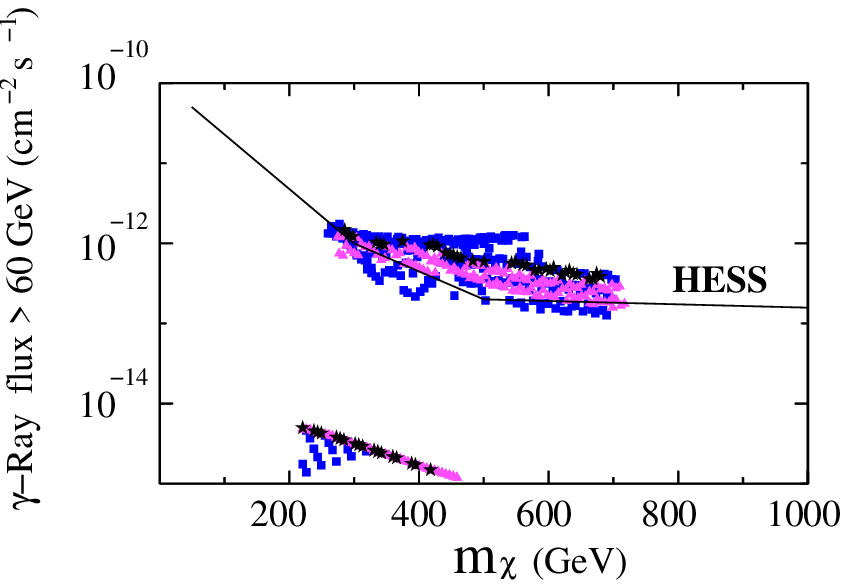,width=0.55\textwidth}
       }
          \caption{{\footnotesize 
The same as in 
Fig.~\ref{fig:univtb5flux} but for $\tan\beta=35$ and 
the non-universal scalar case {\it c)} 
$\delta_1=-1$, $\delta_2=1$, discussed in Eq.~(\ref{3cases}).
}}
        \label{fig:nonuniv3tb5flux}
    \end{center}
\end{figure}

\begin{figure}
\vskip -3cm \hskip 15cm

    \begin{center}
\centerline{
       \epsfig{file=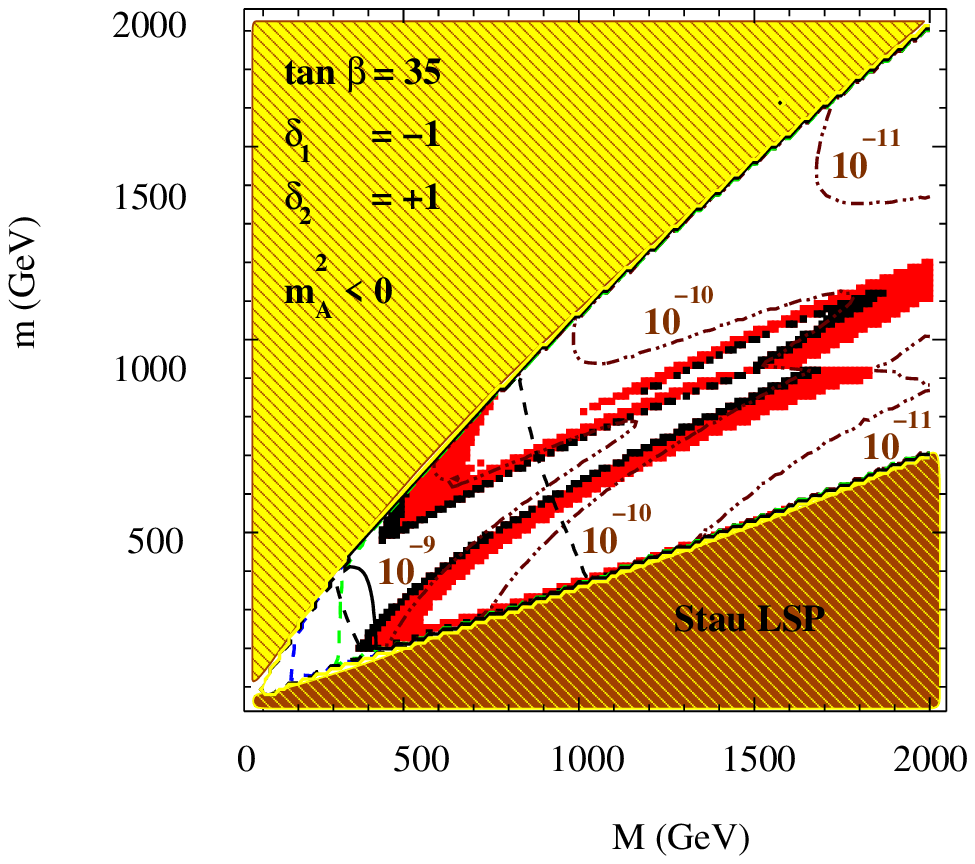,width=0.6\textwidth}
       }
          \caption{{\footnotesize 
The same as in 
Fig.~\ref{fig:univtb5scan} but for $\tan\beta=35$ and 
the non-universal scalar case {\it c)} 
$\delta_1=-1$, $\delta_2=1$, discussed in Eq.~(\ref{3cases}).
The region to the left of the solid line is excluded by $b\to s\gamma$.
 }}
        \label{fig:nonuniv3tb35scan}
    \end{center}
    \begin{center}
\centerline{
       \epsfig{file=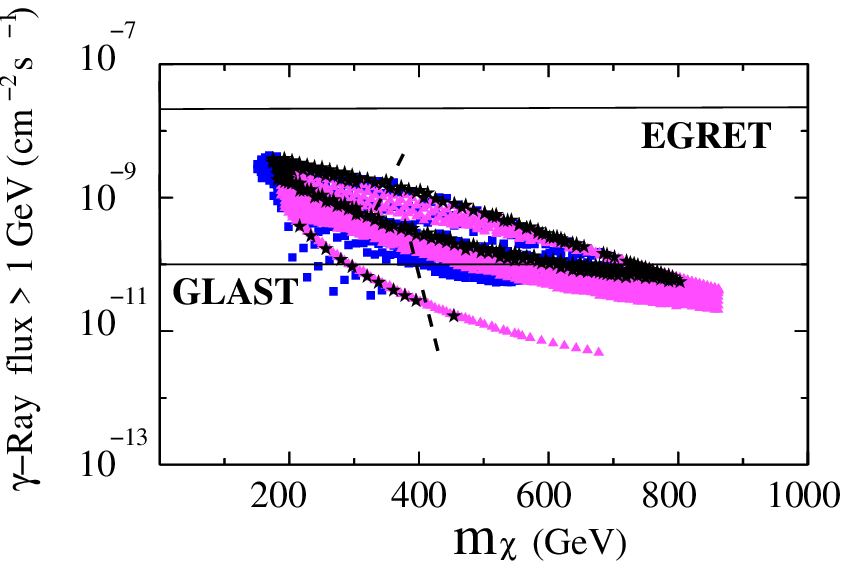,width=0.55\textwidth}\hskip 0.75cm
       \epsfig{file=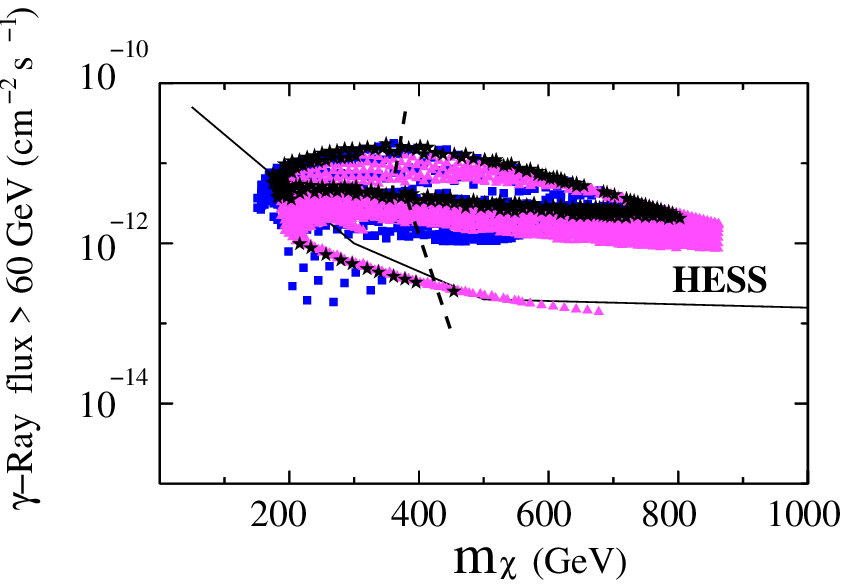,width=0.55\textwidth}
       }
          \caption{{\footnotesize 
The same as in 
Fig.~\ref{fig:univtb5flux} but for $\tan\beta=35$ and 
the non-universal scalar case {\it c)} 
$\delta_1=-1$, $\delta_2=1$, discussed in Eq.~(\ref{3cases}).
Now, the regions to the right of the dashed lines
correspond to
$\asusy< 7.1\times10^{-10}$, and would be excluded by
$e^+e^-$ data.
These do not include the points with black starts between both dashed
lines, which are allowed.
 }}
        \label{fig:nonuniv3tb35flux}
    \end{center}
\end{figure}

Concerning case {\it c)}, this is basically a combination of the previous
two cases, {\it a)} and {\it b)}. 
In Fig.~\ref{fig:nonuniv3tb5scan}
one can see that
the region where the neutralino is mainly Higgsino 
is already open at tan$\beta=5$. This is of course due to the non-universality of $m_{H_u}$.
Consequently, large fluxes can be obtained, as shown in 
Fig.~\ref{fig:nonuniv3tb5flux}.
These figures are qualitatively similar to those corresponding to 
case {\it a)} with tan$\beta=5$, 
Figs.~\ref{fig:nonuniv1tb5scan} and \ref{fig:nonuniv1tb5flux}. 
Moreover, one can see in Fig.~\ref{fig:nonuniv3tb35scan} and \ref{fig:nonuniv3tb35flux} 
for tan$\beta=35$ 
that the $A$ contribution is now also largely 
present, although the effect of the 
Higgsino component of the neutralino is also important in the region
$m, M~\sim$ 500 GeV.
Notice that the shape of the allowed corridors is qualitatively
similar
to that  corresponding to case {\it b)},
Figs.~\ref{fig:nonuniv2tb35scan} and \ref{fig:nonuniv2tb35flux},

 
Let us also mention that
the combination of both non-universalities gives rise to very light
Higgses, particularly charged Higgses, and that new final states ($H^+ W^-$ or
$Z H$) can be kinematically allowed and even reach 30\% of the branching ratio
in some regions of the parameter space.
This effect is not possible in mSUGRA.

 
Finally, we have also analyzed very large values of
tan$\beta$, such as 50, for cases 
{\it a)}, {\it b)}, and {\it c)}.
The results are qualitatively similar, in the sense that
many points accessible for GLAST and HESS are obtained.

\newpage

\subsection{Non-universal gauginos}

Let us now study the effect of the non-universality in the gaugino
masses. We can parameterise this as follows:
\begin{eqnarray}
  M_1=M\ , \quad M_2=M(1+ \delta'_{2})\ ,
  \quad M_3=M(1+ \delta'_{3})
  \ ,
  \label{gauginospara}
\end{eqnarray}
where $M_{1,2,3}$ are the bino, wino and gluino masses, respectively,
and $\delta'_i=0$ corresponds to the universal case.

In order to increase the annihilation cross section
it is worth noticing that
$M_3$ appears in the RGEs of squark masses.
Thus the contribution of squark masses
proportional
to the top Yukawa coupling in the RGE of $m_{H_u}^2$ will
do this less negative if $M_3$ is 
small.
As discussed above, this produces a decrease in the value of
$|\mu|$
and therefore a mixed Higgsino-gaugino lightest neutralino,
which may help in order to increase the gamma-ray flux.

 Because the mass of the lightest Higgs is very dependent on the value
 of $M_3$, through radiative corrections, its decrease is very limited. 
 In fact, in order
 to satisfy the lower limit of $M_3$, $M$ in (\ref{gauginospara}) may have
 to increase, thus rather than a decrease in $M_3$ what
 one obtains is an effective increase in $M_1$ and $M_2$,
 which
 leads to a larger (less
 negative) value of $\higgsu$ and thus a reduction in the value of
 $|\mu|$.
 This in turn
 implies heavier neutralinos, 
 when the lightest neutralino is mostly
 gaugino, and an increase of the Higgsino composition, which would be
 dominant if $M_1>|\mu|$ at the electroweak scale.
 Finally, decreasing the ratio $M_3/M_1$ leads to 
 a more efficient
 neutralino annihilation, 
 due to the enhancement in the Higgsino
 components\footnote{Note that the value of $\higgsd$ also increases,
 thus $m_A$ calculated from (\ref{ma2}) is typically
 not very affected. Although we will see below that the
decrease of $m_A$ in some regions
 can be important for the analysis.}
 of $\neut$, 
 entailing a reduction of $\Omega_{\tilde{\chi}^0_1}$.

\begin{figure}
\vskip -3cm \hskip 15cm

    \begin{center}
\centerline{\epsfig{file=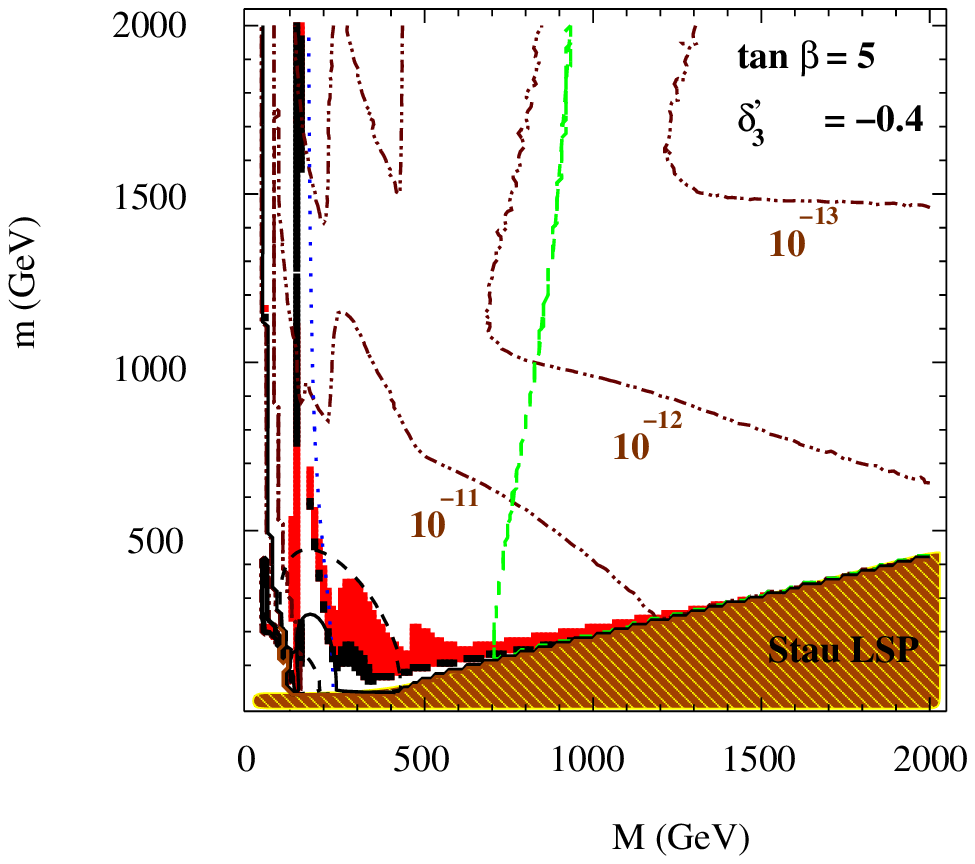,width=0.6\textwidth}}

 
          \caption{{\footnotesize 
The same as in 
Fig.~\ref{fig:univtb5scan} but for  
the non-universal gaugino case 
$\delta_3'=-0.4$, discussed in Eq.~(\ref{gauginospara}).
}}
        \label{fig:nonuniv4tb5scan}
    \end{center}
    \begin{center}
\centerline{
       \epsfig{file=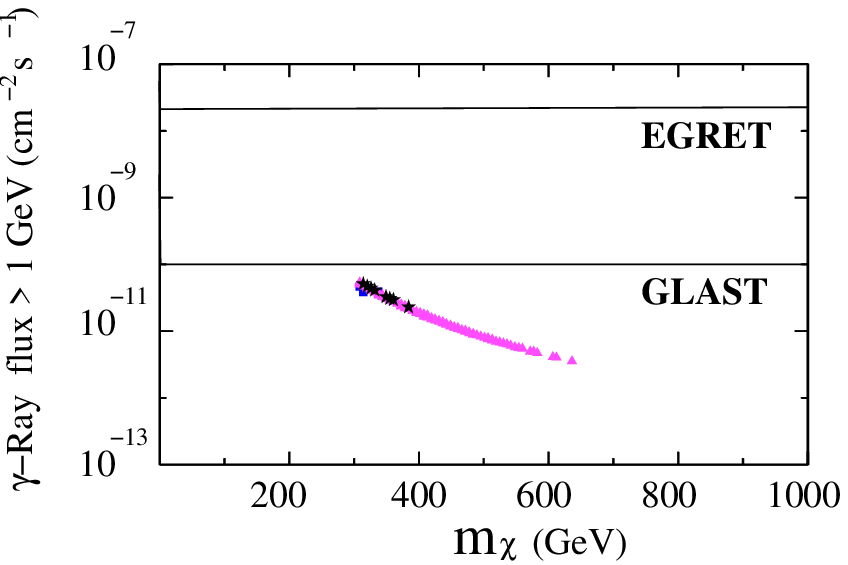,width=0.55\textwidth}\hskip 0.75cm
       \epsfig{file=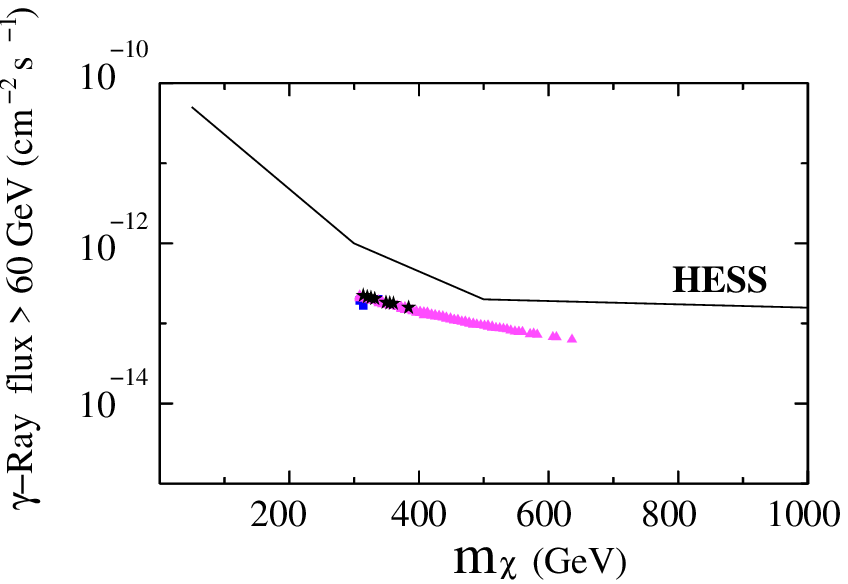,width=0.55\textwidth}
       }
          \caption{{\footnotesize 
The same as in 
Fig.~\ref{fig:univtb5flux} but for  
the non-universal gaugino case 
$\delta_3'=-0.4$, discussed in Eq.~(\ref{gauginospara}).
}}
        \label{fig:nonuniv4tb5flux}
    \end{center}
\end{figure}

In the following we analyze the case
$\delta'_{2}=0, \delta'_{3}=-0.4$. Lower values of $M_3$ at the GUT scale
cannot produce a sufficient amount of relic density.
We show the situation for 
$\tan\beta=5$ in Fig.~\ref{fig:nonuniv4tb5scan}.
Notice that the Higgs mass bound
allows only 
the region corresponding to the narrow coannihilation branch.
Thus the $A$-exchange channel is the relevant one here.
The fact that in this case the gamma-ray flux is larger than in mSUGRA
(one can compare Figs.~\ref{fig:nonuniv4tb5flux} and \ref{fig:univtb5flux})
is due to the different composition of the neutralino.
For example, for a point with 
$m=$ 200 GeV and $M=$ 1100 GeV 
(corresponding to $\Omega_{\chi}$=0.16) one obtains 
$Z_{13}=$ 0.11 in this non-universal case,
whereas  $Z_{13}=$ 0.04 in mSUGRA. 
Thus the Higgsino component of the
neutralino is more important with non-universality.

\begin{figure}
\vskip -3cm \hskip 15cm

    \begin{center}
\centerline{
       \epsfig{file=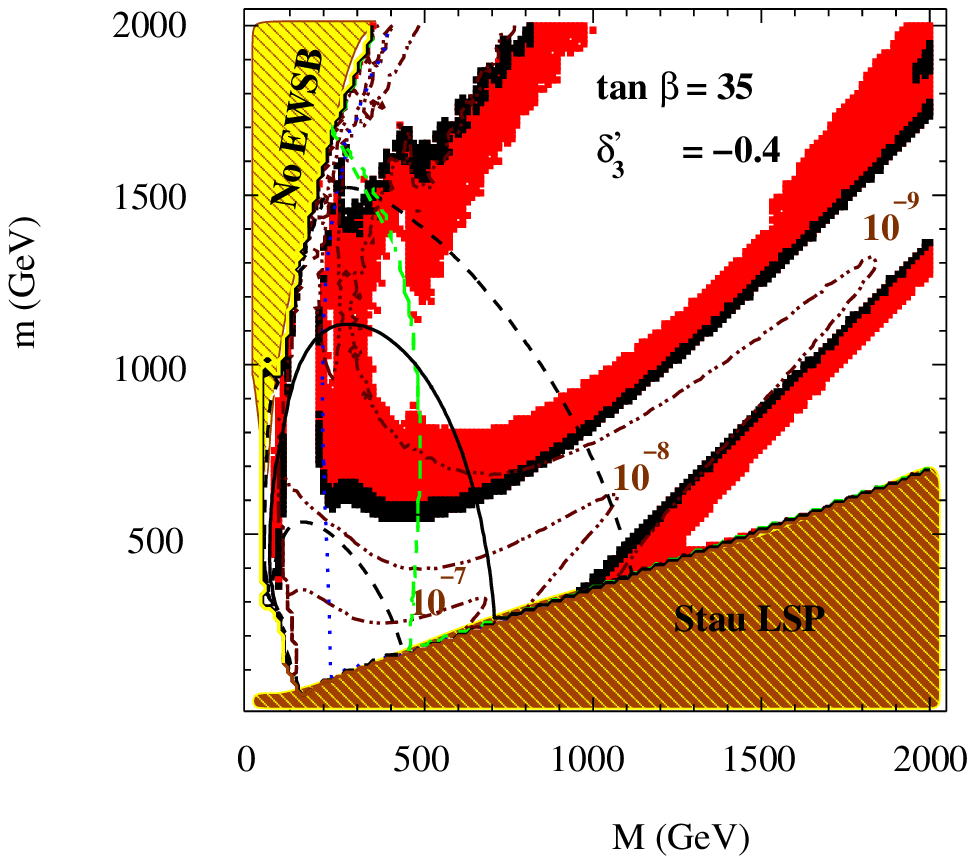,width=0.6\textwidth}
       }
          \caption{{\footnotesize 
The same as in 
Fig.~\ref{fig:univtb5scan} but for
$\tan\beta=35$ and  
the non-universal gaugino case 
$\delta_3'=-0.4$, discussed in Eq.~(\ref{gauginospara}).
The region to the left of the solid line is excluded by $b\to s\gamma$.
}}
        \label{fig:nonuniv4tb35scan}
    \end{center}
\centerline{
       \epsfig{file=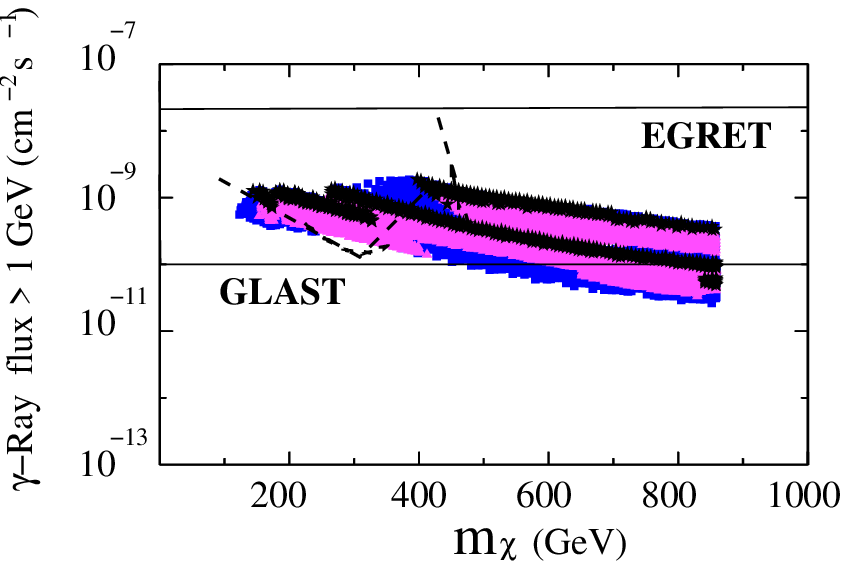,width=0.55\textwidth}\hskip 1cm
       \epsfig{file=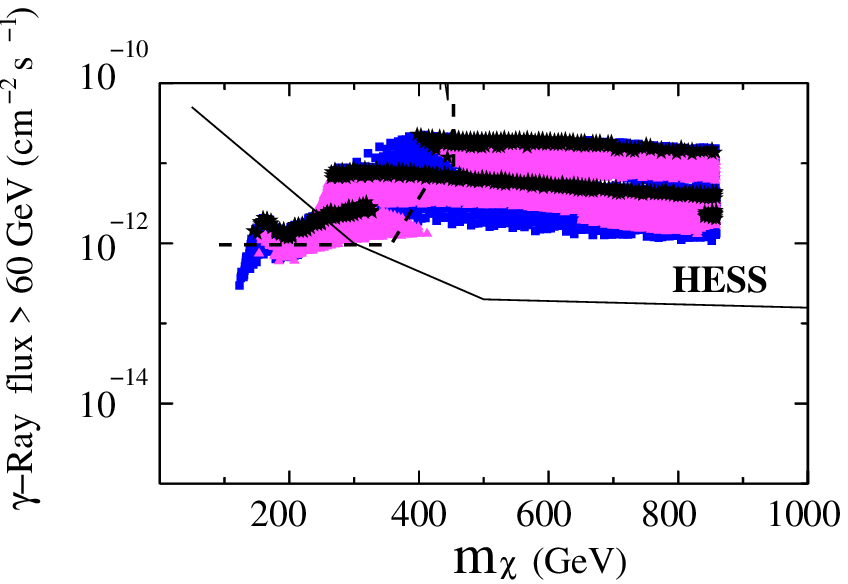,width=0.55\textwidth}
       }
          \caption{{\footnotesize 
The same as in 
Fig.~\ref{fig:univtb5flux} but for
$\tan\beta=35$ and  
the non-universal gaugino case 
$\delta_3'=-0.4$, discussed in Eq.~(\ref{gauginospara}).
Now, the regions to the right (and below) 
the dashed lines 
correspond to
$\asusy< 7.1\times10^{-10}$, and would be excluded by
$e^+e^-$ data.
These do not include the points with black stars between both dashed
lines, which are allowed.
}}
        \label{fig:nonuniv4tb35flux}
\end{figure}

In the large tan $\beta$ regime 
there is now a new feature.
We can see in Fig.~\ref{fig:nonuniv4tb35scan} 
that we have three cosmologically allowed corridors for $\tan\beta=35$.
To analyze them, let us consider a fixed value of $m$, for example
$m=1000$ GeV, 
and go from large to small $M$.
For $M=2000$ GeV, 
$m_A$ is already smaller than 
$2m_{\tilde{\chi}^0_1}$ (this effect is not present for $\tan\beta=5$), producing 
a too large relic density. 
However, for smaller values of $M$ the decreasing in $m_A$ is not 
as steep as in non-universal scalar cases, and for
$M\sim 1500$ GeV,  $m_A$
becomes very close to $2m_{\tilde{\chi}^0_1}$. 
Thus 
the relic density is placed inside the bounds,
$0.094<\Omega_{\tilde{\chi}^0_1}h^2<0.3$.
For slightly smaller values of $M$ one arrives to the A-pole region since
$m_A \sim 2m_{\tilde{\chi}^0_1} $, and
the relic density is too small. If we continue to the left,
$m_A > 2m_{\tilde{\chi}^0_1}$, and this close the A-pole region 
producing a decrease in the
annihilation cross section, which allows us to
enter in the second corridor. 
Finally, the third corridor to the left of
Fig.~\ref{fig:nonuniv4tb35scan} 
is the usual Higgsino branch generated by the increase of the Higgsino
components
of the neutralino, near the noEWSB area where $\mu$ is small.
As discussed in the case of Fig.~\ref{fig:nonuniv1tb35scan},
the gap between this branch and the no EWSB area
increases with larger values of the parameters.
See e.g. $m$=1500 GeV and 2000 GeV.
It is also worth noticing that
now, for a fixed value of $m$, there are more points
with a relic density inside the astrophysical bounds.
This is so because the value of $\mu $ is more stable
with respect to variations in $M$. Note in this respect
that here we are lowering  $\mu $ through the value
of $M_3$, and this is a loop effect.

Let us finally mention that in the latter region
the
$Z$-exchange channel dominates producing fluxes with
$b\overline{b}$ final states
($t\overline{t}$ being kinematically disfavoured). This corresponds to $M$ smaller than
about 900 GeV.
The corridors related to the A-pole open for 
$M$ about 600 and 1000 GeV.
These three corridors can be clearly distinguished in Fig.~\ref{fig:nonuniv4tb35flux}.
The mixed Higgsino-gaugino one for $200\lsim m_{\chi}\lsim  400$ GeV, 
and the other two
for $300 \lsim m_{\chi} \lsim 900$ GeV and 
$450\lsim m_{\chi}\lsim 900$ GeV. The latter gives rise to 
the largest flux.
As in the case of non-universal scalars, we can observe in the figure
that important regions of the parameter space will be accessible for
future experiments.



\newpage

\subsection{General case: non-universal scalars and gauginos}



 



Let us now consider the general case where the soft
supersymmetry-breaking terms for both scalar and
gauginos have a non-universal structure.
This was studied in detail for the case of direct detection
in Ref.~\cite{cggm03-12}, and we will follow closely the analysis. 
We will be mostly interested 
in analyzing possible departures with respect to the parameter space
which was allowed in the previos subsections, producing large values
of
the neutralino annihilation cross section.
For this reason, we
will concentrate on the interesting choices for scalar
non-universalities given in Eq.~(\ref{3cases}), 
and study the effect of adding gaugino non-universalities to these.

We will consider 
the possibility of increasing the value
of $M_3$ with respect to $M_1$, which can be done with
$\delta'_3>0$ in Eq.~(\ref{gauginospara}).
In this case, 
the constraint on the Higgs mass and on \bsg\ 
will be satisfied for smaller
values of $M$, and therefore the effective value of $M_1$ can be
smaller than in the universal case.
Thus lighter neutralinos can be obtained.

\begin{figure}
\vskip -3cm \hskip 15cm

    \begin{center}
\centerline{
       \epsfig{file=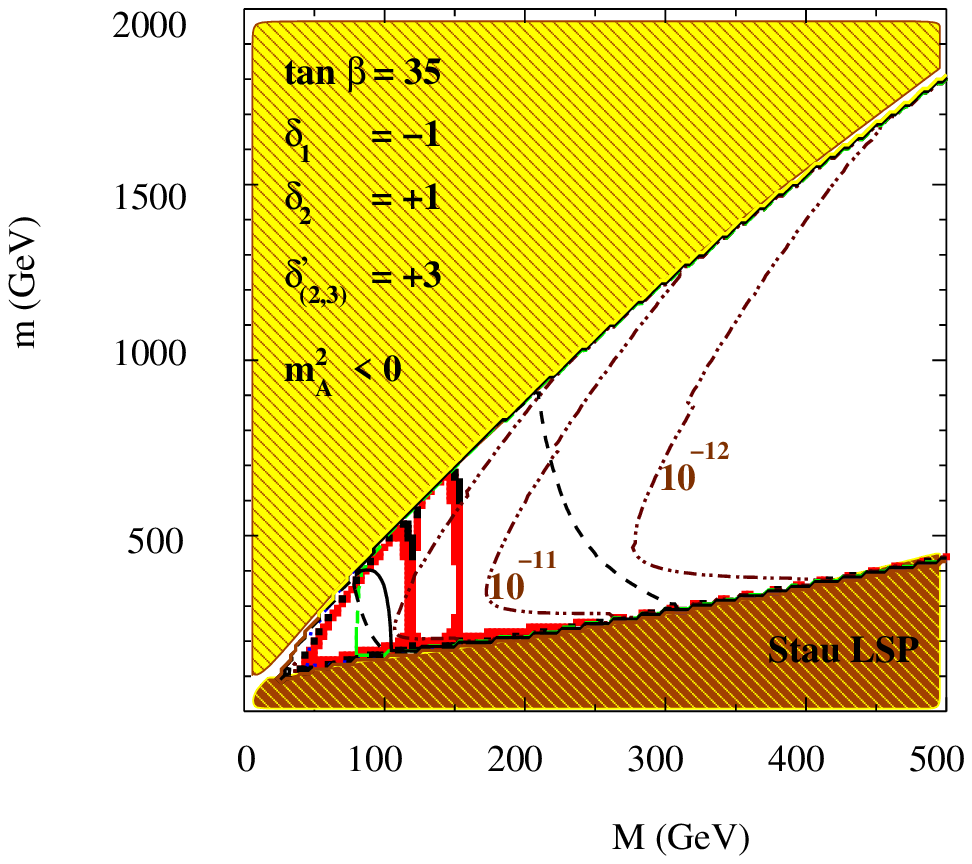,width=0.6\textwidth}
       }
          \caption{{\footnotesize 
The same as in 
Fig.~\ref{fig:univtb5scan} but for $\tan\beta=$35 and 
the non-universal scalar and gaugino case
$\delta_2=1, \delta_1=-1$,
$\delta_2'=\delta_3'=3$.
The region to the left of the solid line is excluded by $b\to s\gamma$.
}}
        \label{fig:nonuniv10ctb35scan}
    \end{center}
\centerline{
       \epsfig{file=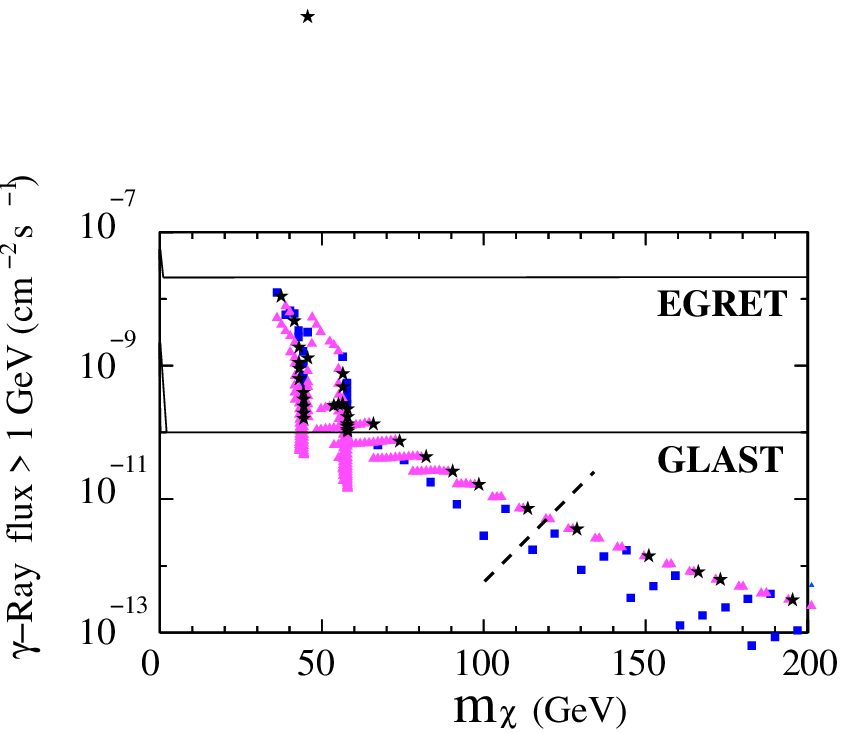,width=0.55\textwidth}\hskip 1cm
       \epsfig{file=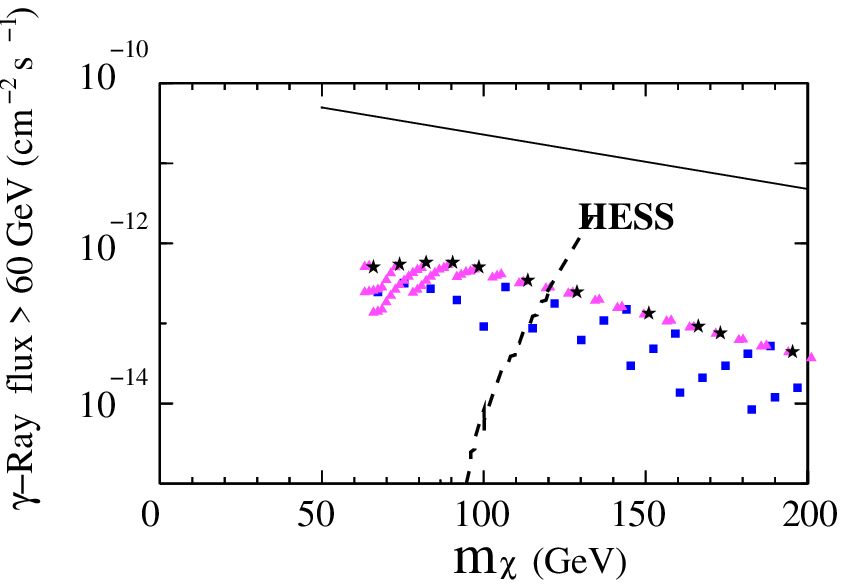,width=0.55\textwidth}
       }
          \caption{{\footnotesize 
The same as in 
Fig.~\ref{fig:univtb5flux} but for $\tan\beta=$35 and 
the non-universal  scalar and gaugino case
$\delta_2=1, \delta_1=-1$,
$\delta_2'=\delta_3'=3$,
using the same parameter space as in 
Fig.~\ref{fig:nonuniv10ctb35scan}.
Now, the region to the right of the 
dashed line
corresponds to
$\asusy< 7.1\times10^{-10}$, and would be excluded by
$e^+e^-$ data.
}}
        \label{fig:nonuniv10ctb35flux}
\end{figure}

In particular, we will study 
the theoretical predictions for the gamma-ray flux 
for an example with $\delta'_{2,3}=3$, $\tan\beta=35$, $A=0$, and the
three choices of
Higgs non-universalities of Eq.~(\ref{3cases}).
As we will see, this choice of gaugino parameters
favours the appearance of light neutralinos within the GLAST
sensitivity range. 
In these cases the typical values for the relic density
are too large and therefore not consistent with the WMAP result. 
A very effective decrease can be achieved when
the $h$ and $Z$-poles are crossed at $\neumass=m_h/2$ and $M_Z/2$,
respectively, giving rise to a very effective neutralino
annihilation through the corresponding $s$-channels. This 
is evidenced by the narrow
chimneys in the cosmologically preferred regions in
Fig.~\ref{fig:nonuniv10ctb35scan},
where case {\it c)} of 
Eq.~(\ref{3cases}) is shown.
Because of the
very efficient decrease in the CP-odd Higgs mass in this case, annihilation of very
light neutralinos can be boosted and thus the correct relic density
obtained. This allows the
existence of 
neutralinos with $\neumass\sim30$ GeV which produce a flux 
even close to EGRET sensitivity, as shown in Fig.~\ref{fig:nonuniv10ctb35flux}. 
For the other two cases, {\it a)} and {\it b)} of Eq.~(\ref{3cases}), the figures are
qualitatively
similar, although the fluxes are slightly smaller, bounded by
$10^{-9}$ cm$^{-2}$ s$^{-1}$.
For the three cases these light neutralinos
can in principle be
detected by GLAST. 
Of course,
they  cannot be detected by 
atmospheric Cherenkov telescopes, as e.g. HESS, 
since these have a large threshold energy of about 50 GeV.


It is worth noticing that the existence of very light staus with
light neutralinos in case {\it a)} with 
$\delta'_{2,3}=3$,
makes it possible the domination of the 
lightest stau exchange channel with respect to 
the $A$ one (see Fig.~\ref{fig:feynmandetail}) in regions of the
parameter
space, producing a large amount of taus.
This comes from the RGE governing the $m_{L_3}$ evolution.
Indeed, decreasing $M_1$ for a given $M_2$ decreases the value of $m_{L_3}$
at the electroweak scale, giving a lighter stau. 
On the other hand, cases 
$b)$ and $c)$ do not exhibit such a behavior
because 
decreasing
at the same time $m_{H_d}$ cancels the effect (the two contributions appear
with an opposite sign in the RGE). 

Let us now discuss the possibility of obtaining even lighter neutralinos.
One of the requirements for the appearance of such very low
neutralinos is to have $M_1\ll\mu, M_2$ at low energy (thus having
almost pure binos). This can be achieved with adequate choices of$b\overline{b}$
gaugino non-universalities, in particular with $\delta'_{2,3}\gg1$. 
However, as mentioned above,
without a very effective reduction of $m_A$, the relic density would
be too large, and therefore inconsistent with observations. 
Here the presence of non-universal scalars is
crucial. In particular, non-universalities as the ones we have
described in the Higgs sector in
Eq.~(\ref{3cases}) provide a very effective way of lowering $m_A$ and are
thus optimal for this purpose.

More specifically, it is in case  {\it b)} and especially  {\it c)} where the
reduction in 
$m_A$ is more effective (not being so constrained by regions with
$\mu^2<0$) and for this reason very light neutralinos 
can easily appear. We have already seen in the previous example
how this happens
for case  {\it c)} with $\tan\beta=35$ and $\delta'_{2,3}=3$ (see
Fig.~(\ref{fig:nonuniv10ctb35flux})). 
On the contrary, in case  {\it a)}  higher values of $\tan\beta$ are required in
order to further reduce the value of $m_A$. 
One can check explicitly that 
$\tan\beta\gsim33$ is sufficient to obtain $\neumass<M_Z/2$ in cases
{\it b)} and  {\it c)}, whereas $\tan\beta\gsim45$ is necessary in case {\it a)}. In all
the three cases $\delta'_{2,3}\gsim3$ leads to these results.

\begin{figure}
\vskip -3cm \hskip 15cm

    \begin{center}
\centerline{
       \epsfig{file=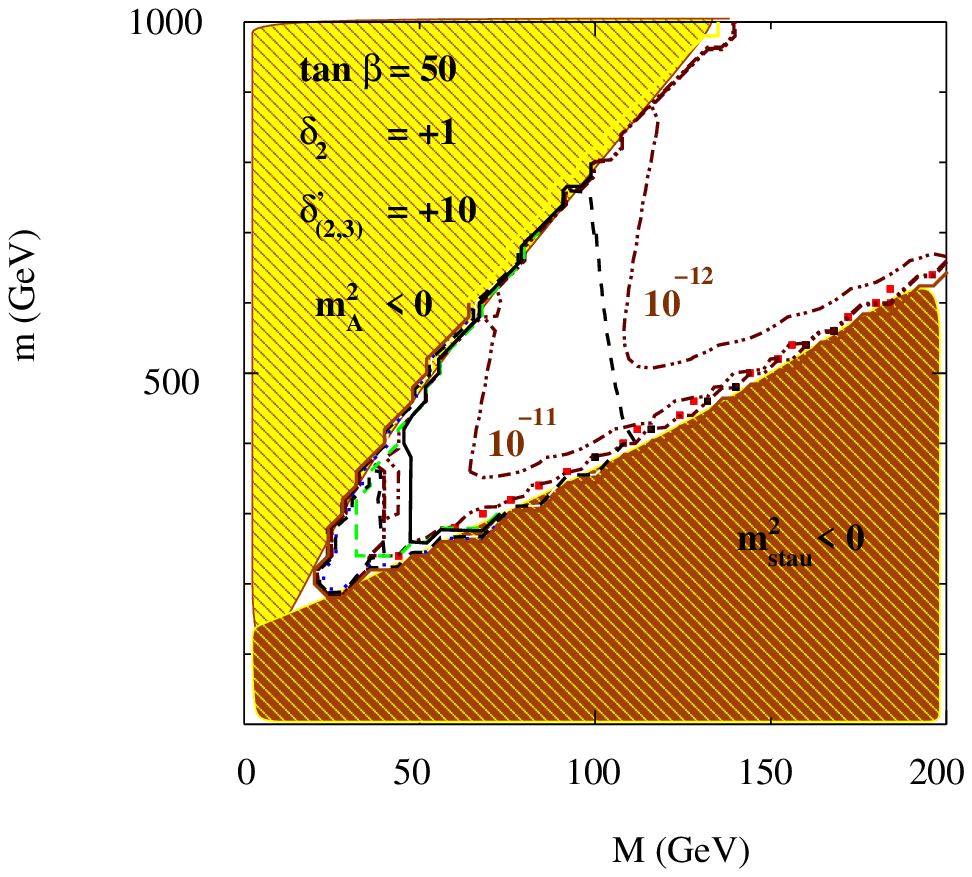,width=0.6\textwidth}
       }
          \caption{{\footnotesize 
The same as in 
Fig.~\ref{fig:univtb5scan} but for $\tan\beta=$50 and 
the non-universal  scalar and gaugino case
$\delta_2=1, \delta_1=0$,
$\delta_2'=\delta_3'=10$.
The region to the left of the solid line is excluded by $b\to s\gamma$.
}}
        \label{fig:nonuniv11atb50scan}
    \end{center}
\centerline{
       \epsfig{file=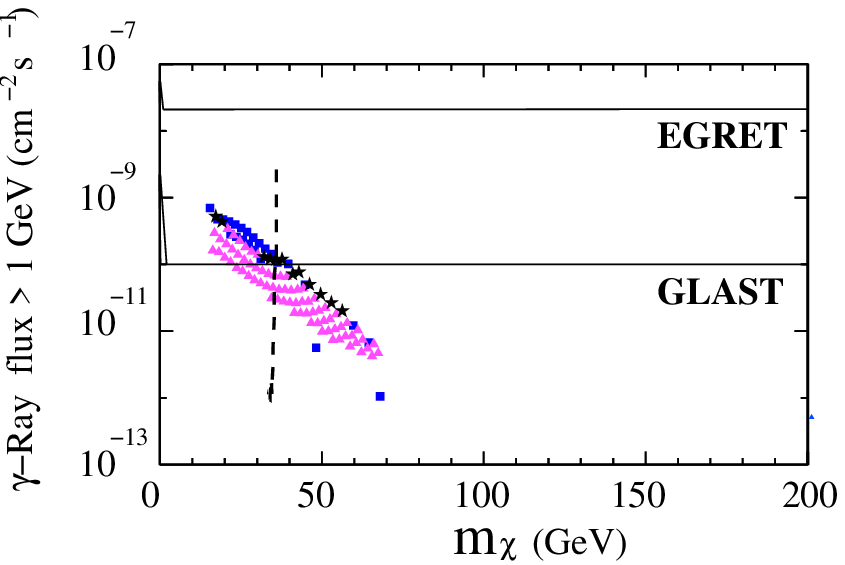,width=0.55\textwidth}\hskip 1cm
       \epsfig{file=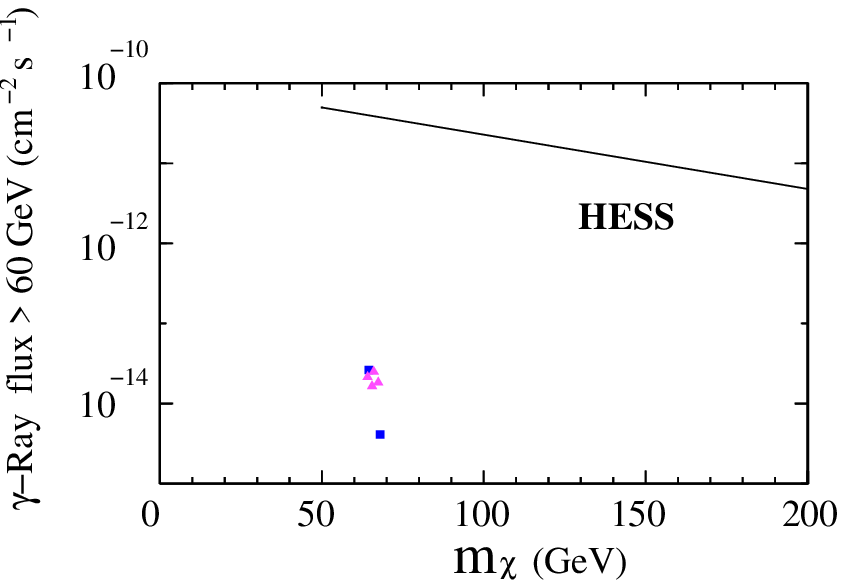,width=0.55\textwidth}
       }
          \caption{{\footnotesize 
The same as in 
Fig.~\ref{fig:univtb5flux} but for $\tan\beta=50$ and 
the non-universal scalar and gaugino case
$\delta_2=1, \delta_1=0$,
$\delta_2'=\delta_3'=10$,
using the same parameter space as in
Fig.~\ref{fig:nonuniv11atb50scan}. Now, in the left frame,
the region to the right of the 
 dashed line
corresponds to
$\asusy< 7.1\times10^{-10}$, and would be excluded by
$e^+e^-$ data. In the right frame all points would be excluded by
this constraint.
}}
        \label{fig:nonuniv11atb50flux}
\end{figure}

Let us therefore complete our discussion 
by analysing the case $\delta'_{2,3}=10$ for 
the choice  {\it a)}  of
scalar non-universalities in Eq.~(\ref{3cases}). The result
is shown in Fig.~\ref{fig:nonuniv11atb50flux}
for  $\tan\beta=50$.
We observe
the appearance of very light neutralinos, 
whose
cross section can be in
range of detectability of near-future experiments.
In particular,
points with a gamma-ray flux in the region of GLAST sensitivity
are obtained with
$\neumass\sim 15$ GeV.
As mentioned for $\delta'_{2,3}=3$, 
these points  cannot be detected by HESS because of the 
large threshold energy of about 50 GeV.
For the other two cases, {\it b)} and {\it c)} of Eq.~(\ref{3cases}), the figures are
qualitatively similar. 

The effect of the different constraints on the
corresponding $(m,M)$ parameter space is represented in
Fig.~\ref{fig:nonuniv11atb50scan}.
Note that due to the large value of $\delta'_{3}$
the
resonances with the lightest Higgs and the $Z$ 
are not sufficient
to place the relic abundance inside the bounds, and
the regions giving rise to very light
neutralinos with a consistent relic density are extremely narrow.
In these points the mass of the CP-odd Higgs can be very close to its
experimental limit, $m_A\lsim100$ GeV. 
Since the neutralino mass is so small, the region
excluded due to the neutralino not being the LSP is
negligible. However, now the lower bound on the stau mass
plays an important role. In fact,
an important region of the parameter space
is excluded for having $m_{\tilde{\tau}_1}^2<0$.

\newpage



\section{Conclusions}

We have analyzed the theoretical predictions for 
the indirect detection of neutralino dark matter
in the context of a general SUGRA theory,
where both the scalar and gaugino soft supersymmetry-breaking terms
can have a non-universal structure.
In particular, we have computed the predictions for the gamma-ray
fluxes arising from the galactic center due to neutralino annihilation,
and compared it with the sensitivity of detectors.

Although the presence of non-universal scalar and gaugino masses
is not able to explain the 
intriguing signal detected by the space-based experiment EGRET, 
the annihilation cross section increases significantly
with respect to the universal (mSUGRA) case,
producing larger gamma-ray fluxes.
For example, 
whereas mSUGRA is not compatible 
for $\tan\beta<35$ with the sensitivity of another
crucial projected experiment, GLAST,
non-universality in the scalar sector
can increase the gamma-ray fluxes about two orders of magnitude,
entering in the region which will be analyzed by this experiment.
For the atmospheric Cherenkov detector HESS, which has already
begun operations, the situation is qualitatively similar.
Of course, for larger values of $\tan\beta$ a similar
increase in the gamma-ray fluxes is obtained. 
The above effects are mainly due to the Higgsino nature of the
neutralino and/or the decrease of the CP-odd Higgs ($A$) mass
in some regions of the parameter space with non-universalities.

When non-universal gauginos are considered, 
one can also obtain regions of the parameter space 
with annihilation cross sections much larger than in 
mSUGRA.
This is also the case when it is combined
with non-universal scalars, i.e. the most general case in SUGRA.
In addition, in this general case, 
departures from the allowed parameter space obtained
in the previous situations are possible.
For example, neutralinos 
as light as
$\neumass\sim 15$ GeV, and with large fluxes
can be obtained.

This general analysis can be very useful in the study of more specific
cases, such as the SUGRA theories resulting at the low energy
limit of string constructions. In particular,
D-brane scenarios in type I string theory give rise to theories
where non-universalities appear both in the scalar and gaugino sectors.

Let us finally remark that the above results have been obtained
using the standard NFW density profile for our galaxy.
However, we have discussed how they can easily be
extended for other widely used profiles,
such as Isothermal, Kravtsov et al., Navarro et al., and 
Moore et al.
In particular, one can use Table~\ref{tab} where the function
$\bar J(\Delta \Omega)$ has been computed for the different profiles
(and possible angular resolutions). This simply enters
as a scaling factor in the gamma-ray fluxes.
Obviously, a Moore et al. profile, which is about two orders of 
magnitude larger than the NFW one, would modify drastically
the results concerning EGRET.

\vspace*{1cm}
\noindent{\bf Acknowledgements}

Y. Mambrini thanks E. Nezri and D.G. Cerde\~no
for valuable discussions and A.M. Teixeira for the help provided during 
this study.
The work of Y. Mambrini was supported 
by the European Union under contract
HPRN-CT-2000-00148.
The work of C. Mu\~noz was supported 
in part by the Spanish DGI of the
MEC 
under contracts BFM2003-01266 and FPA2003-04597;
and the European Union under contract 
HPRN-CT-2000-00148.

\nocite{}
\bibliography{bmn}
\bibliographystyle{unsrt}

\end{document}